\documentclass[10pt,aps,pra,twocolumn,nofootinbib]{revtex4-1}   

\usepackage[dvips]{epsfig}
\usepackage{amsmath}    
\usepackage{amsfonts}  
\usepackage{amssymb}
\usepackage{graphicx}   
\usepackage{mathrsfs}
\usepackage{ulem}
\usepackage{soul}
\usepackage[usenames,dvipsnames]{color}

\usepackage{xcolor}

\usepackage[colorlinks]{hyperref}

\usepackage{braket}

\usepackage{colortbl}

\definecolor{gray(x11gray)}{rgb}{0.75, 0.75, 0.75}

\definecolor{darksienna}{rgb}{0.24, 0.08, 0.08}

\newcommand{\Cov}{\textrm{Cov}}

\newcommand{\SymCov}{\widehat{\text{Cov}}}

\usepackage{xcolor}

\newcommand{\red}{\color{red}}

\AtBeginDocument{%
    \hypersetup{
      urlcolor     = blue, 
      filecolor   = red,
      linkcolor    = blue, 
      citecolor   = blue, 
    }
} 

\begin{document}

\title{Optimal phase sensitivity of an unbalanced Mach-Zehnder interferometer}

\author{Karunesh~K.~Mishra$^{1,2}$, Stefan~Ataman$^1$}
\affiliation{$^1$Extreme Light Infrastructure - Nuclear Physics (ELI-NP), `Horia Hulubei' National R\&D Institute for Physics and Nuclear Engineering (IFIN-HH), 30 Reactorului Street, 077125 M\u{a}gurele, jud. Ilfov, Romania}
\email{stefan.ataman@eli-np.ro}
\affiliation{$^2$Department of Physics, Institute of Science, Banaras Hindu Univesity, Varanasi-221005, India}

\date{\today}


\begin{abstract}
In this paper we address the problem of optimizing an unbalanced Mach-Zehnder interferometer, for a given pure input state and considering a specific detection scheme. While the optimum transmission coefficient of the first beam splitter can be uniquely determined via the quantum Fisher information only [Phys. Rev. A \textbf{105}, 012604 (2022)], the second beam splitter transmission coefficient is detection-scheme dependent, too. We systematically give analytic solutions for the optimum transmission coefficient of the second beam splitter for three types of widely used detection schemes. We provide detailed examples including both Gaussian and non-Gaussian input states, showing when an unbalanced Mach-Zehnder interferometer can outperform its balanced counterpart in terms of phase sensitivity.

\end{abstract}

\maketitle

\section{Introduction}
\label{sec:introduction}


Optical interferometers, alongside with their atomic counterparts, are among the most precise metrological devices when it comes to detection of small signals. It is no accident that LIGO \cite{LIGO2018} and Virgo \cite{Ace14} are laser-based interferometers. Quantum enhanced metrology \cite{Gio04,Cav81} promises significant precision improvements in different areas of quantum technologies \cite{Acin2018}, gravitational wave astronomy \cite{LIGO13,Tse2019}, quantum-enhanced dark matter searches \cite{Backes2021} and biological samples measurements \cite{Tay13,Taylor2016}.
The engineering of strong squeezed vacuum states of light is a key technology for the reduction of quantum noise \cite{Meh18}. 


By using classical resources one can arrive at the so-called shot-noise limit, $ \Delta \varphi_{\mathrm{SNL}} \sim 1/\sqrt{\bar{N}}$, where $ \bar{N}$  denotes the average number of input photons \cite{Par09}. Employing quantum resources, such as squeezed \cite{Cav81,Pez08,Pezze2013,Andersen2016}, NOON \cite{Hol93} or other non-classical \cite{Birrittella2012,Kim1998} states one can approach the ultimate quantum limit $\Delta \varphi_{\mathrm{HL}} \sim 1/\bar{N} $ also called as the Heisenberg limit (HL) \cite{Giovannetti2006}.


In this work we focus on the phase sensitivity of a Mach-Zehnder interferometer (MZI), however, as it is well known \cite{Dem15,Pezze2013}, most interferometers can be mapped into a MZI.


Achieving the theoretically best phase sensitivity for an interferometer is an important goal, since one would like to optimize over all possible estimators and for all possible detection schemes. The answer to this problem is given by the 
quantum Fisher information (QFI) \cite{Helstrom1967,Helstrom1968,Holevo1973,Wu19}. The QFI's ($\mathcal{F}$) pivotal importance stems from its connection to the quantum Cram\'er-Rao bound (QCRB), \emph{i. e.} $\Delta\varphi_{QCRB}=1/\sqrt{\mathcal{F}}$. Thus, the phase sensitivity of any practical detection scheme is bound to be $\Delta\varphi_{det}\geq\Delta\varphi_{QCRB}$.


However, it was realized that by employing the above (single-parameter QFI) definition leads to an over-estimation of the available performance \cite{Jar12}. It was thus realized that the role of a potentially available (even if not explicitly used) external phase reference \cite{Jar12,Dem15} should not be neglected. Proper ways to discard resources that are actually unavailable have been put forward \cite{Jar12}. These include input state phase averaging and the introduction of the two-parameter QFI \cite{Jar12,Lan13}. However, an external phase reference is often available, and thus the aforementioned single-parameter QFI yields an attainable performance \cite{Ata20}.


The phase sensitivity of a balanced Mach-Zehnder interferometer (MZI) depends on a number of parameters, including the input state \citep{API18,Ata19,Ata20,Luo2022,Pezze2006,Shin1999} and the  employed detection scheme \cite{DAriano1994,Zho17}. 

Among the proposed input states we mention the large class of Gaussian states which includes the coherent plus squeezed vacuum \cite{Cav81,Pez08}, squeezed coherent plus squeezed vacuum \cite{Par95,Ata19} and squeezed coherent plus squeezed coherent \cite{Spa15,Ata19} input states. When it applies, we also discuss the important role of the input phase-matching conditions (PMC) \cite{Liu13,API18,Ata19}.

When the input state is non-Gaussian, aside from the well-known NOON \cite{Hol93} states, the coherent plus Fock input \cite{Birrittella2012} has been shown to have a quantum metrological interest.


When considering an unbalanced MZI \cite{Zhong2020,Ata20,Tak17}, the transmission coefficients of the two BS come also into play. As discussed in reference \cite{Ata22}, the first BS can be unequivocally optimized via the QFI. 
Although the balanced (50:50) MZI scenario yields the optimal two-parameter QFI a large class of input states, this is no longer true when an external phase reference available \cite{Ata20}.


Many detection schemes have been reported in the literature \cite{Gar17,Yuen1983,Ramakrishnan2022,Birrittella2021,Zhong2021,Campos2003} including difference intensity (also called direct detection) \citep{API18,Zho17}, single mode \cite{API18,Zho17,Gar17}, balanced homodyne \cite{Yuen1983,Gar17} and  parity \cite{Campos2003,Birrittella2021} detection.


In this paper we address the optimization in terms of the phase sensitivity for an unbalanced MZI. The input state is assumed to be pure and the optimization is carried out for each considered detection scheme. As mentioned previously, the transmission coefficient of the first BS can be unambiguosly optimized with the help of QFI \cite{Ata22}. We thus thoroughly address the second beam splitter's transmission coefficient and the working point's optimization for each considered detection scheme. We give analytical solutions for all the optima involved. We are also able to explain many previously results reported in the literature.

Examples are given for a number of Gaussian \cite{Gar17,Ata19,Par95} and non-Gaussian \cite{Birrittella2012} input states, outlining situations when an unbalanced MZI can outperform its balanced counterpart in terms of phase sensitivity.



This paper is structured as follows. In Section \ref{sec:interferometric_setup} we describe our interferometric setup, make some notation conventions, introduce the quantum Fisher information and provide the quantum Cram\'er-Rao bound. In Section \ref{sec:detection_schemes} we characterize in detail the three considered detection schemes.  In Section \ref{sec:phase_sens_opt_general} we describe the general phase sensitivity optimization procedure in the sense of minimizing the phase uncertainty $\Delta\varphi$. We thoroughly discuss via examples our results in Section \ref{sec:detector_performance}. Finally, conclusions are drawn in Section \ref{sec:conclusions}.

\section{Interferometric setup}
\label{sec:interferometric_setup}
\begin{figure}
\includegraphics[width=0.48\textwidth]{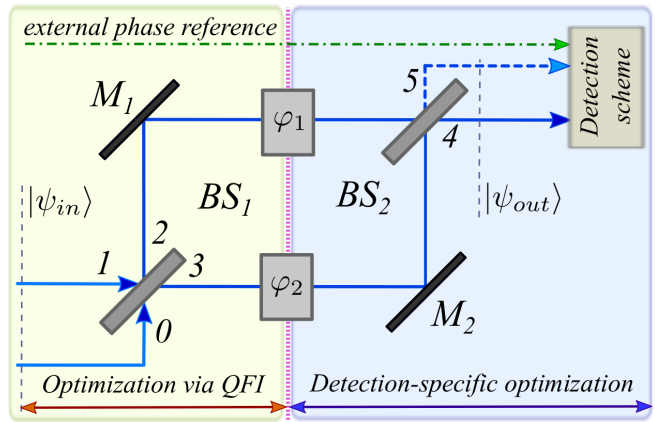}
\caption{The interferometric experimental setup considered in this work. The transmission coefficient ($T$) of the first beam splitter is optimized via the QFI, as discussed in reference \cite{Ata22}. The transmission coefficient of the second BS ($T'$) is optimized by taking into account the specific detection scheme employed.}
\label{fig:Fig1}
\end{figure}

We consider the Mach-Zehnder interferometric setup depicted in Fig.~\ref{fig:Fig1}. The input state is assumed pure and the interferometer is characterized by the two beam splitters having transmission coefficients $T$ (for $BS_1$), respectively, $T'$ (for $BS_2$). We consider the most general scenario encompassing two internal phase shifts, $\varphi_1$ ($\varphi_2$) in the upper (lower) arm of the interferometer. 

Energy conservation for a symmetrical (\emph{i. e.} thin-film) beam splitter  imposes the usual constraints ${|T|^2+|R|^2=1}$ and $T^*R+TR^*=0$ \cite{GerryKnight}. Here $R$ denotes the reflection coefficient of the first beam splitter ($BS_1$). Throughout this work we will use the parametrization
\begin{equation}
\label{eq:BS1_vartheta_parametrization}
\left\{
\begin{array}{l}
T=\cos\frac{\vartheta}{2}\\
R=i\sin\frac{\vartheta}{2}
\end{array}
\right.
\end{equation}
and similarly for $BS_2$,
\begin{equation}
\label{eq:BS2_vartheta_parametrization}
\left\{
\begin{array}{l}
T'=\cos\frac{\vartheta'}{2}\\
R'=i\sin\frac{\vartheta'}{2},
\end{array}
\right.
\end{equation}
where  $R'$ denotes the reflection coefficient of the second beam splitter and $\{\vartheta,\vartheta'\}\in[0,\pi]$. We introduce the input Schwinger pseudo-angular momentum operators \cite{Yurke1986,Campos1989},
\begin{equation}
\label{eq:QO_BS_and_MZI:Schwinger_J1_operator}
\hat{J}_x=\frac{\hat{a}_0^\dagger\hat{a}_1+\hat{a}_0\hat{a}_1^\dagger}{2},
\end{equation}
\begin{equation}
\label{eq:QO_BS_and_MZI:Schwinger_J2_operator}
\hat{J}_y=\frac{\hat{a}_0^\dagger\hat{a}_1-\hat{a}_0\hat{a}_1^\dagger}{2i},
\end{equation}
and
\begin{equation}
\label{eq:QO_BS_and_MZI:Schwinger_J3_operator}
\hat{J}_z=\frac{\hat{a}_0^\dagger\hat{a}_0-\hat{a}_1^\dagger\hat{a}_1}{2},
\end{equation}
where $\hat{a}_l$ ($\hat{a}_l^\dagger$) denote the usual annihilation (creation) operators for the input modes $l=0,1$ \cite{GerryKnight}. We also introduce the input total photon number operator,
\begin{equation}
\label{eq:N_is_a_dagger_a_0_plus_1}
\hat{N}
=\hat{n}_0+\hat{n}_1,
\end{equation}
where $\hat{n}_l=\hat{a}_l^\dagger\hat{a}_l$ denotes the usual number operator for a mode $l$. 

\subsection{First BS optimization via QFI}

We encounter two main scenarios in this optimization and we briefly outline them below. In the case when no external phase reference in available, the only relevant phase information is
\begin{equation}
\varphi=\varphi_2-\varphi_1,
\end{equation}
and one needs to use the so-called two-parameter QFI in order to discharge resources that are unavailable. We thus introduce the two-parameter quantum Fisher information \cite{Jar12,Ata19,Ata20,Ata22} (see details in Appendix \ref{sec:app:single_two_parameter_QFI}),
\begin{equation}
\label{eq:Fisher_information_F_2p_DEFINITION}
\mathcal{F}^{(2p)}
=\mathcal{F}_{dd}-\frac{\mathcal{F}_{sd}^2}{\mathcal{F}_{ss}},
\end{equation}
and this QFI implies the QCRB,
\begin{equation}
\label{eq:Delta_varphi_QCRB_2p_DEFINTION}
\Delta\varphi_{QCRB}^{(2p)}=\frac{1}{\sqrt{\mathcal{F}^{(2p)}}}.
\end{equation}
This limit will be relevant especially for the difference-intensity (direct) detection scheme described in Section \ref{subsec:diff_intensity_mode} and for the single-mode intensity detection scheme described in Section \ref{subsec:single_mode}.

The optimum transmission coefficient $T^{(2p)}_{opt}$ of the first beam splitter that maximizes the two-parameter QFI \eqref{eq:Fisher_information_F_2p_DEFINITION} is found in reference \cite{Ata22}, Section VA. In all performance plots from Section \ref{sec:detector_performance} involving a detection scheme having no access to an external reference, we will assume for the first BS the transmission coefficient $T^{(2p)}_{opt}$, thus in all future calculations we will consider ${\vartheta = 2\arccos T^{(2p)}_{opt}}$.

If an external phase reference is available and assuming the internal phase shift convention \cite{Ata22},
\begin{equation}
\label{eq:one_phase_shift_CONVENTION}
\left\{
\begin{array}{l}
\varphi_1=0\\
\varphi_2=\varphi,
\end{array}
\right.
\end{equation}
the (asymmetric) single-parameter QFI \cite{Jar12,Ata20,Ata22},
\begin{equation}
\label{eq:F_i_is_four_times_variance_n3}
\mathcal{F}^{(i)}=4\Delta^2{\hat{n}_3}
\end{equation}
must be employed (see also Appendix \ref{sec:app:single_two_parameter_QFI}). The single-parameter QFI $\mathcal{F}^{(i)}$ implies the QCRB
\begin{equation}
\label{eq:Delta_varphi_QCRB_i_DEFINTION}
\Delta\varphi^{(i)}_{QCRB}=\frac{1}{\sqrt{\mathcal{F}^{(i)}}}.
\end{equation}
The optimum transmission coefficient $T^{(i)}_{opt}$ of the first beam splitter that maximizes the asymmetric single-parameter QFI \eqref{eq:F_i_is_four_times_variance_n3} is given in  reference \cite{Ata22}, Section VB. All performance plots from Section \ref{sec:detector_performance} involving a balanced homodyne detection (BHD) scheme will use the convention from Eq.~\eqref{eq:one_phase_shift_CONVENTION} and also assume that $BS_1$ is characterized by $T^{(i)}_{opt}$, thus in all calculations we will use ${\vartheta = 2\arccos T^{(i)}_{opt}}$.

We mention that a sub-case to the scenario with an external phase reference is possible, namely by replacing Eq.~\eqref{eq:one_phase_shift_CONVENTION} with the convention
\begin{equation}
\label{eq:two_phases_CONVENTION}
\left\{
\begin{array}{l}
\varphi_1=-\frac{\varphi}{2}\\
\varphi_2=\frac{\varphi}{2}.
\end{array}
\right.
\end{equation}
As discussed in the literature \cite{Jar12,Ata20,Ata22}, this $\pm\varphi/2$ scenario is described by the QFI $\mathcal{F}^{(ii)}=\Delta^2({\hat{n}_2}-{\hat{n}_3})$ \cite{Jar12} that can also be expressed in respect with the two-parameter Fisher matrix elements as $\mathcal{F}^{(ii)}=\mathcal{F}_{dd}$ \cite{Ata22}. This QFI implies the QCRB $\Delta\varphi^{(ii)}_{QCRB}={1}/{\sqrt{\mathcal{F}^{(ii)}}}$. Since this $\pm\varphi/2$  convention presents little interest for an unbalanced MZI scenario, we will not consider it in this work.

\subsection{Field operator transformations for an unbalanced MZI}
We have the input-output field operator transformations,
\begin{equation}
\label{eq:field_op_transf_a4_a5_vs_a0_a1}
\left\{
\begin{array}{l}
\hat{a}_4={\mathcal{A}}_{40}{\hat{a}_0}+{\mathcal{A}}_{41}{\hat{a}_1}\\
\hat{a}_5={\mathcal{A}}_{50}{\hat{a}_0}
+{\mathcal{A}}_{51}{\hat{a}_1},
\end{array}
\right.
\end{equation}
where the $\mathcal{A}$-coefficients are given by (see \emph{e. g.} \cite{Ata14,Ata18c})
\begin{equation}
\label{eq:A_coeffs_def_two_phases}
\left\{
\begin{array}{l}
{\mathcal{A}}_{40}=TT'e^{-i\varphi_1}+RR'e^{-i\varphi_2}\\
{\mathcal{A}}_{41}=TR'e^{-i\varphi_2}+RT'e^{-i\varphi_1}\\
{\mathcal{A}}_{50}=TR'e^{-i\varphi_1}+RT'e^{-i\varphi_2}\\
{\mathcal{A}}_{51}=TT'e^{-i\varphi_2}+RR'e^{-i\varphi_1}.
\end{array}
\right.
\end{equation}
Employing the conventions \eqref{eq:BS1_vartheta_parametrization}-\eqref{eq:BS2_vartheta_parametrization} and assuming a single internal phase shift \eqref{eq:one_phase_shift_CONVENTION},
Eq.~\eqref{eq:A_coeffs_def_two_phases} can be rewritten as
\begin{equation}
\label{eq:A_coeffs_one_phase_angles}
\left\{
\begin{array}{l}
{\mathcal{A}}_{40} = \cos\frac{\vartheta}{2}\cos\frac{\vartheta'}{2}
-\sin\frac{\vartheta}{2}\sin\frac{\vartheta'}{2}e^{-i\varphi}\\
{\mathcal{A}}_{41} = i\left(\cos\frac{\vartheta}{2}\sin\frac{\vartheta'}{2}e^{-i\varphi}
+\sin\frac{\vartheta}{2}\cos\frac{\vartheta'}{2}\right)\\
{\mathcal{A}}_{50} = i\left(\cos\frac{\vartheta}{2}\sin\frac{\vartheta'}{2}
+\sin\frac{\vartheta}{2}\cos\frac{\vartheta'}{2}e^{-i\varphi}\right)\\
{\mathcal{A}}_{51}=\cos\frac{\vartheta}{2}\cos\frac{\vartheta'}{2}e^{-i\varphi}
-\sin\frac{\vartheta}{2}\sin\frac{\vartheta'}{2}.
\end{array}
\right.
\end{equation}
For future convenience, we introduce the following $K$-coefficients:
\begin{equation} 
\label{eq:Kx_Ky_Kz_trigonometric}
\left\{
\begin{array}{l}
{ K_x} = {\sin\vartheta'}{\sin{\varphi}}\\
{ K_y} = -\left(\sin\vartheta{\cos\vartheta'}
+\cos\vartheta{\sin\vartheta'}{\cos{\varphi}}\right)\\
{ K_z} = \cos\vartheta{\cos\vartheta'}-\sin\vartheta\sin\vartheta'{\cos{\varphi}}
\end{array}
\right.
\end{equation}
and the above terms obey the constraint
\begin{equation}
\label{eq:Kx_Ky_Kz_square_sum_is_1}
{ K_x^2}+{ K_y^2}+{ K_z^2}=1.
\end{equation}
Also, by direct calculation we get the following results connecting the $K$- and $\mathcal{A}$-coefficients:
\begin{equation}
\label{eq:Kd_Zd_constraints_A_coeffs}
\left\{
\begin{array}{l}
|{\mathcal{A}}_{40}|^2=|{\mathcal{A}}_{51}|^2=\frac{1}{2}\left(1+K_z\right)\\
|{\mathcal{A}}_{50}|^2=|{\mathcal{A}}_{41}|^2=\frac{1}{2}\left(1-K_z\right)\\
\Re\left\{{\mathcal{A}}_{40}{\mathcal{A}}^*_{41}\right\}
=-\Re\left\{{\mathcal{A}}_{50}{\mathcal{A}}^*_{51}\right\}
=\frac{ K_{x}}{2}\\
\Im\left\{{\mathcal{A}}_{40}{\mathcal{A}}^*_{41}\right\}
=-\Im\left\{{\mathcal{A}}_{50}{\mathcal{A}}^*_{51}\right\}
=\frac{ K_{y}}{2},
\end{array}
\right.
\end{equation}
where $\Re$ ($\Im$) denotes the real (imaginary) part.

\section{Detection schemes}
\label{sec:detection_schemes}
For a general detection scheme employing an observable $\hat{O}(\varphi)$, the phase sensitivity can be defined via the error propagation formula \cite{GerryKnight,Dem15}
\begin{equation}
\label{eq:Delta_varphi_DEF}
\Delta\varphi=\frac{\sqrt{\Delta^2\hat{O}(\varphi)}}
{\Big\vert\frac{\partial\braket{\hat{O}(\varphi)}}{\partial\varphi}\Big\vert},
\end{equation}
where the variance of the operator $\hat{O}(\varphi)$ is defined by
\begin{equation}
\label{eq:Variance_O_DEF}
\Delta^2\hat{O}=\braket{\hat{O}(\varphi)^2}-\braket{\hat{O}(\varphi)}^2.
\end{equation}
For clarity, from this point on, we will not write explicitly the operator's $\varphi$-dependence, \emph{i. e.} we will write $\hat{O}$ instead of $\hat{O}(\varphi)$. 

In the following, we will determine the phase sensitivity corresponding to each of the three considered detection schemes.

\begin{figure}
\includegraphics[width=0.49\textwidth]{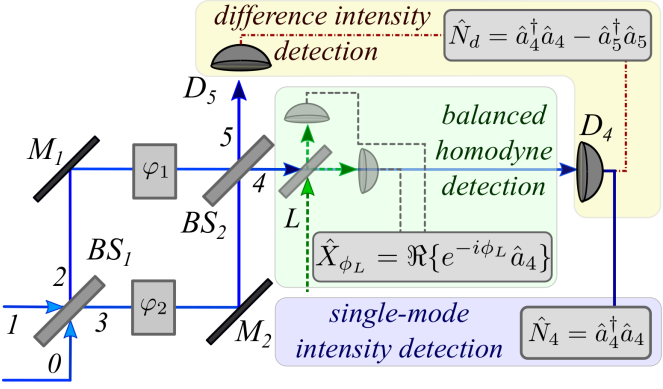}
\caption{The three detection schemes considered in this work. For the difference-intensity (direct) detection scheme (see Sec.~\ref{subsec:diff_intensity_mode}) we have the output operator $\hat{N}_d=\hat{n}_4-\hat{n}_5$. For the single-mode intensity detection scheme we use $\hat{n}_4$ as output operator (see Sec.~\ref{subsec:single_mode}), while for the balanced homodyne detection scheme (see Sec.~\ref{subsec:BHD_scheme}) we employ $\hat{X}_L$ given in Eq.~\eqref{eq:X_average_homodyne_versus_a4}.}
\label{fig:Fig2}
\end{figure}

\subsection{Difference intensity detection scheme}
\label{subsec:diff_intensity_mode}

In the difference-intensity detection scheme, we employ as observable the difference between the output photo-currents, $\hat{N}_d=\hat{n}_4-\hat{n}_5$ (see Fig.~\ref{fig:Fig2}). Expressing $\hat{N}_d$ in respect with the input field operators yields
\begin{equation}
\label{eq:Nd_fct_Nd_Re-Im_Zd_Jx_Jy_Jz}
\hat{N}_d=2K_x\hat{J}_x+2K_y\hat{J}_y+2K_z\hat{J}_z
\end{equation}
and we have
\begin{eqnarray}
\label{eq:del_avg_Nd_del_varphi_fct_Nd_Re-Im_Zd_Jx_Jy_Jz}
\frac{\partial\braket{\hat{N}_d}}{\partial\varphi}
=2\left(\braket{\hat{J}_x}\cos{\varphi}
+(\cos\vartheta\braket{\hat{J}_y}
\right.
\nonumber\\
\left.
+\sin\vartheta\braket{\hat{J}_z})\sin{\varphi}
\right)\sin\vartheta'.
\end{eqnarray}
After some calculations (see Appendix \ref{sec:app:diff_calculations}), the variance of the operator $\hat{N}_d$ is found to be 
\begin{eqnarray}
\label{eq:Nd_Variance_Kx_Ky_Kz_Jx_Jy_Jz}
\Delta^2{\hat{N}_d} =
4K_x^2\Delta^2{\hat{J}_x}
+4K_y^2\Delta^2{\hat{J}_y}
+4K_z^2\Delta^2{\hat{J}_z}
\nonumber\\
+8K_xK_z\SymCov\left({\hat{J}_x},{\hat{J}_z}\right)
+8K_xK_y\SymCov\left({\hat{J}_x},{\hat{J}_y}\right)
\nonumber\\
+8K_yK_z\SymCov\left({\hat{J}_y},{\hat{J}_z}\right),
\end{eqnarray}
where the symmetrized covariance of two non-commuting operators $\hat{A}$ and $\hat{B}$ is defined by
\begin{equation}
\label{eq:SymmetrizedCovariance_DEF}
\SymCov\left({\hat{A}},\hat{B}\right)=\frac{\braket{\hat{A}\hat{B}}+\braket{\hat{B}\hat{A}}}{2}-\braket{\hat{A}}\braket{\hat{B}}.
\end{equation}
From definition \eqref{eq:Variance_O_DEF} and using the previous results, the phase sensitivity for a difference-intensity detection scheme is
\begin{eqnarray}
\label{eq:Delta_varphi_diff_generic}
\Delta\varphi_{df}=\frac{1}{2|\sin\vartheta'|}\times
\nonumber\\
\times\frac{\sqrt{\Delta^2{\hat{N}_{d}}}}
{\Big\vert\braket{\hat{J}_x}\cos{\varphi}
+\left(\cos\vartheta\braket{\hat{J}_y}
+\sin\vartheta{\braket{\hat{J}_z}}\right)\sin{\varphi}\Big\vert}.
\end{eqnarray}

\subsection{Single mode intensity detection scheme}
\label{subsec:single_mode}
In this scenario we consider a single photo-current at the output port $4$, the observable conveying information is thus $\hat{n}_4$ (see Fig.~\ref{fig:Fig2}). Expressing it in respect with the input field operator yields
\begin{equation}
\label{eq:n4_fct_Nd_Re-Im_Zd_Jx_Jy_Jz}
{\hat{n}_4}
=\frac{1}{2}{\hat{N}}
+K_x{\hat{J}_x}
+K_y{\hat{J}_y}
+K_z{\hat{J}_z}
\end{equation}
and thus
\begin{eqnarray}
\label{eq:del_n4_avg__del_varphi_fct_Jx_Jy_Jz}
\frac{\partial\braket{\hat{n}_4}}{\partial\varphi}
=\left(\braket{\hat{J}_x}\cos{\varphi}
+\cos\vartheta\sin{\varphi}\braket{\hat{J}_y}
\right.
\nonumber\\
\left.
+\sin\vartheta\sin{\varphi}{\braket{\hat{J}_z}}\right)\sin\vartheta'.
\end{eqnarray}
The variance is found to be
(see details in Appendix \ref{sec:app:single_calculations})
\begin{eqnarray}
\label{eq:n4_Variance_fct_Kd_Zd_J_Schwinger}
\Delta^2{\hat{n}_4}
=\frac{1}{4}\Delta^2{\hat{N}}
+K_x^2\Delta^2{\hat{J}_x}+K_y^2\Delta^2{\hat{J}_y}
+K_z^2\Delta^2{\hat{J}_z}
\nonumber\\
+2K_xK_y\SymCov\left({\hat{J}_x},{\hat{J}_y}\right)
+2K_xK_z\SymCov\left({\hat{J}_x},{\hat{J}_z}\right)
\nonumber\\
+2K_yK_z\SymCov\left({\hat{J}_y},{\hat{J}_z}\right)
+K_x\Cov\left({\hat{J}_x},{\hat{N}}\right)
\nonumber\\
+K_y\Cov\left({\hat{J}_y},{\hat{N}}\right)
+K_z\Cov\left({\hat{J}_z},{\hat{N}}\right),
\end{eqnarray}
where the covariance of two operators $\hat{A}$ and $\hat{B}$ is defined as usually by
\begin{equation}
\label{eq:Covariance_DEF}
\Cov\left({\hat{A}},\hat{B}\right)=\braket{\hat{A}\hat{B}}-\braket{\hat{A}}\braket{\hat{B}}.
\end{equation}
The phase sensitivity \eqref{eq:Delta_varphi_DEF} for this scenario is
\begin{eqnarray}
\label{eq:Delta_varphi_sing_generic}
\Delta\varphi_{sg}=\frac{1}{\vert\sin\vartheta'\vert}\times
\nonumber\\
\times\frac{\sqrt{\Delta^2{\hat{n}_4}}}
{\Big\vert\braket{\hat{J}_x}\cos{\varphi}
+\left(\cos\vartheta\braket{\hat{J}_y}
+\sin\vartheta{\braket{\hat{J}_z}}\right)\sin{\varphi}\Big\vert}.
\end{eqnarray}

\subsection{Balanced homodyne detection scheme}
\label{subsec:BHD_scheme}
If we assume a balanced homodyne detection (BHD) scheme at the output port $4$ (see Fig.~\ref{fig:Fig2}), the operator modeling this  detection scheme is
\begin{equation}
\label{eq:X_average_homodyne_versus_a4}
\hat{X}_{\phi_L}
=\frac{e^{-i\phi_L}\hat{a}_4+e^{i\phi_L}\hat{a}_4^\dagger}{2}.
\end{equation}
We find the variance of the above operator,
\begin{eqnarray}
\label{eq:Delta_2_X_versus_a4}
\Delta^2\hat{X}_{\phi_L}
=\frac{1}{4}
+\frac{1}{2}\left(\Cov\left({\hat{a}_4^\dagger},{\hat{a}_4}\right)
+\Re\left\{e^{-i2\phi_L}{\Delta^2\hat{a}_4}\right\}\right),\:\:
\quad
\end{eqnarray}
where by the covariance term above we obviously mean 
\begin{equation}
\Cov\left({\hat{a}_4^\dagger},{\hat{a}_4}\right)=\braket{\hat{n}_4}-|\braket{\hat{a}_4}|^2.
\end{equation}
In the single internal phase shift scenario \eqref{eq:one_phase_shift_CONVENTION}, by
using the field operator transformations \eqref{eq:field_op_transf_a4_a5_vs_a0_a1} and \eqref{eq:A_coeffs_one_phase_angles} we find
\begin{eqnarray}
\label{eq:del_X_del_varphi_single_phase}
\frac{\partial\braket{\hat{X}_{\phi_L}}}{\partial\varphi}
=\bigg(\sin\frac{\vartheta}{2}\Re\left\{ie^{-i(\phi_L+\varphi)}
\braket{\hat{a}_0}\right\}
\nonumber\\
+\cos\frac{\vartheta}{2}
\Re\left\{e^{-i(\phi_L+\varphi)}\braket{\hat{a}_1}\right\}
\bigg)\sin\frac{\vartheta'}{2}.
\end{eqnarray}
The phase sensitivity in this case is found to be
\begin{eqnarray}
\label{eq:Delta_varphi_hom_1phase_generic}
\Delta\varphi_{hom}=\frac{1}{\sin\frac{\vartheta'}{2}}\times
\nonumber\\
\times\frac{\sqrt{\Delta^2{\hat{X}_{\phi_L}}}}{
\vert\sin\frac{\vartheta}{2}\Re\left\{e^{-i(\phi_L+\varphi)}
i\braket{\hat{a}_0}\right\}
+\cos\frac{\vartheta}{2}
\Re\left\{e^{-i(\phi_L+\varphi)}\braket{\hat{a}_1}\right\}
\vert},
\quad\quad
\end{eqnarray}
where $\Delta^2{\hat{X}_{\phi_L}}$ is given in Eq.~\eqref{eq:Delta_2_X_versus_a4}. More details are found in Appendix \ref{sec:app:BHD}.

\begin{figure}
\includegraphics[width=0.49\textwidth]{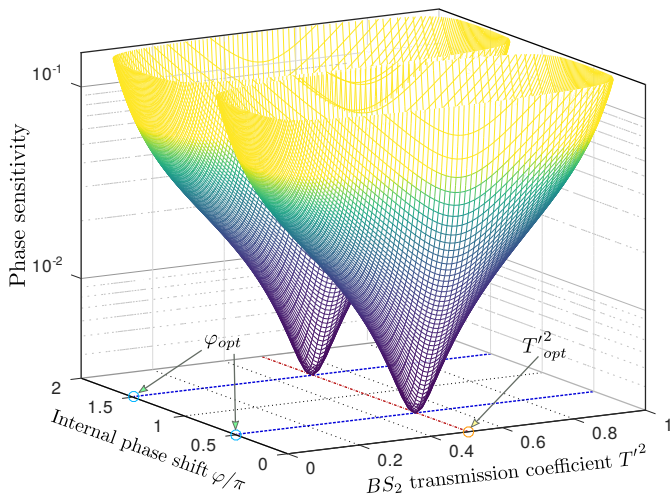}
\caption{Example plot of the phase sensitivity $\Delta\varphi_{df}$ for a difference-intensity detection scheme given in Eq.~\eqref{eq:Delta_varphi_diff_generic} versus the total internal phase shift ($\varphi$) and the $BS_2$ transmission coefficient (${T'}^2$). A coherent plus squeezed vacuum input state Eq.~\eqref{eq:psi_in_coh_plus_sqz_vac} is employed and $BS_1$ is optimized via QFI yielding $T=1/\sqrt{2}$. The optimum in terms of phase sensitivity is found by imposing the working point(s) $\varphi_{opt}=\frac{\pi}{2}+k\pi$ ($k\in\mathbb{Z}$) and $BS_2$ balanced. Parameters used: $\vert\alpha\vert=10^2$, $r=1.2$, and input PMC $2\theta_\alpha-\theta=0$.}
\label{fig:Phase_sens_3D_difference}
\end{figure}

\section{Phase sensitivity optimization}
\label{sec:phase_sens_opt_general}
Since our MZI is unbalanced, we find that our phase sensitivity -- as expected -- depends on the statistics of the input state but also on the transmission coefficients of both BS, plus the two internal phase shifts, $\varphi_1$ and $\varphi_2$. As mentioned previously, we assume the first beam splitter (parametrized by $\vartheta$) to be already optimized via the QFI \cite{Ata22}. This guarantees that up to the phase shifts (see Fig.~\ref{fig:Fig1}) the interferometer is optimized, no matter what beam splitter or detection scheme are employed next. Our phase sensitivity is thus a function of three variables, $\Delta\varphi({\vartheta'},\varphi_1,\varphi_2)$. For a detection scheme not having access to an external phase reference, the phase sensitivity simplifies to $\Delta\varphi({\vartheta'},\varphi)$, where $\varphi=\varphi_2-\varphi_1$. We wish to optimize this phase sensitivity and this is obviously an extremization problem applied to a two variable function \cite{Abramowitz1972}. We thus impose the constraints
\begin{equation}
\label{eq:del_Delta_varphi_del_vartheta_prime_varphi_equal_ZERO}
\left\{
\begin{array}{l}
\partial_{\vartheta'}\Delta\varphi({\vartheta'},\varphi)
=\frac{\partial\Delta\varphi({\vartheta'},\varphi)}{\partial\vartheta'}=0\\
\partial_\varphi\Delta\varphi({\vartheta'},\varphi)
=\frac{\partial\Delta\varphi({\vartheta'},\varphi)}{\partial\varphi}=0.
\end{array}
\right.
\end{equation}
yielding a number of solutions, $({\vartheta'_{opt}},\varphi_{opt})$. However, in the general case, the resulting pairs $({\vartheta'_{opt}},\varphi_{opt})$, are not necessarily extrema points. One must thus also compute the second order derivatives $\partial_{\vartheta'\vartheta'}\Delta\varphi$, $\partial_{\varphi\varphi}\Delta\varphi$, and $\partial_{\vartheta'\varphi}\Delta\varphi$ and impose the constraints
\begin{equation}
\label{eq:vartheta_opt_prime_varphi_opt_local_MINIMA_conditions}
\left\{
\begin{array}{l}
\partial_{\vartheta'\vartheta'}\Delta\varphi({\vartheta'_{opt}},\varphi_{opt})\partial_{\varphi\varphi}
\Delta\varphi({\vartheta'_{opt}},\varphi_{opt})\\
\qquad\qquad\qquad\qquad-\left(\partial_{\vartheta'\varphi}\Delta\varphi({\vartheta'_{opt}},\varphi_{opt})\right)^2>0\\
\partial_{\vartheta'\vartheta'}\Delta\varphi({\vartheta'_{opt}},\varphi_{opt})>0\\
\partial_{\varphi\varphi}\Delta\varphi({\vartheta'_{opt}},\varphi_{opt})>0,
\end{array}
\right.
\end{equation}
in order to find the pair(s) $({\vartheta'_{opt}},\varphi_{opt})$ actually yielding a minimum of the function $\Delta\varphi$.


Nonetheless, the phase sensitivity $\Delta\varphi({\vartheta'},\varphi)$ has usually a well defined minimum\footnote{Of course, degenerate cases (\emph{i. e.} when $T_{opt}=0/1$ and/or $T'_{opt}=0/1$) may appear as result of this optimization. They may also appear while finding the optimum transmission coefficient of $BS_1$ derived from the QFI, as discussed in reference \cite{Ata22}. Quite often, such poor results are the consequence of bad input parameter/PMC choices.} and the optimal solutions are unambiguous. An example is given in Fig.~\ref{fig:Phase_sens_3D_difference}, where we plot the phase sensitivity for a coherent plus squeezed vacuum input given in Eq.~\eqref{eq:psi_in_coh_plus_sqz_vac} and a difference-intensity detection scheme (see Section \ref{subsec:diff_intensity_mode}). Being in a scenario without an external phase reference, optimizing the two-parameter QFI points towards a balanced $BS_1$, \emph{i. e.} $T_{opt}=1/\sqrt{2}$ (or, equivalently $\vartheta_{opt}=\pi/2$) \cite{Ata22}. As detailed in Appendix \ref{subsec:app:Opt_Tprime_diff_mode}, for the input state \eqref{eq:psi_in_coh_plus_sqz_vac} the best performance in phase sensitivity is found when employing a balanced MZI ($T'_{opt}=1/\sqrt{2}$) and a working point $\varphi_{opt}={\pi}/{2}+k\pi$ with $k\in\mathbb{Z}$ (see Fig.~\ref{fig:Phase_sens_3D_difference}). 


For the difference-intensity detection scenario numerous input states will point towards the same optimal settings, among them the squeezed-coherent plus squeezed vacuum \eqref{eq:psi_in_sqzcoh_plus_sqz_vac}, the coherent plus Fock \eqref{eq:psi_in_Fock_plus_coherent} and for most input phase matching conditions, the squeezed-coherent plus squeezed-coherent input (see Appendices \ref{subsec:app:Opt_Tprime_diff_mode} and \ref{subsec:app:Opt_varphi_diff_mode} for a broader discussion).

\begin{figure}
\includegraphics[width=0.49\textwidth]{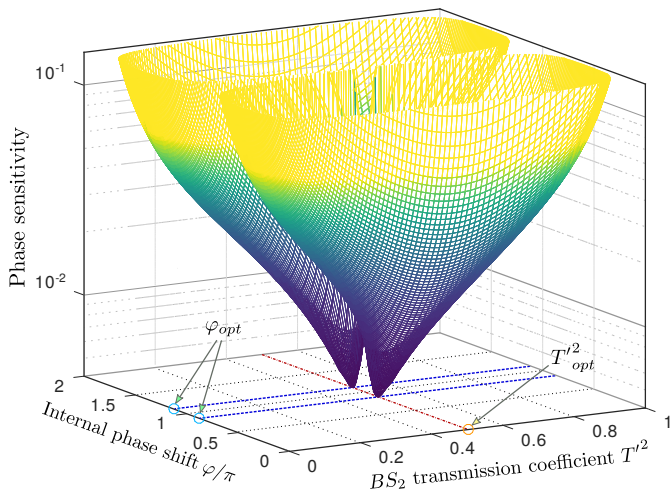}
\caption{Example plot of the phase sensitivity $\Delta\varphi_{sg}$ given in Eq.~\eqref{eq:Delta_varphi_sing_generic} for a single mode intensity detection scheme versus the total internal phase shift ($\varphi$) and the $BS_2$ transmission coefficient (${T'}^2$). A coherent plus squeezed vacuum input state given in Eq.~\eqref{eq:psi_in_coh_plus_sqz_vac} is assumed and $BS_1$ is optimized via the QFI yielding $T=1/\sqrt{2}$. The optimum in terms of phase sensitivity is found for the working point(s) $\varphi_{opt}\approx\pi\pm0.124\pi$ and for $BS_2$ balanced. Parameters used: $\vert\alpha\vert=10^2$, $r=1.2$, and $2\theta_\alpha-\theta=0$.}
\label{fig:Phase_sens_SING_3D_plot}
\end{figure}


Another example is given in Fig.~\ref{fig:Phase_sens_SING_3D_plot}, where we plot the phase sensitivity for a MZI fed by the same input state however using a single-mode intensity detection scheme. Optimizing $BS_1$ via the QFI points towards a balanced solution \emph{i. e.} $T_{opt}=1/\sqrt{2}$ \cite{Ata22}. Similar to the difference-intensity detection scheme, the optimum transmission coefficient of the second beam splitter (see details in Appendix \ref{subsec:app:Opt_Tprime_single}) is found in the balanced case. However, the optimum working point is not found at a multiple of $\pi/2$, as it was previously the case. Indeed, the working point $\varphi_{opt}$ for a single-mode intensity detection is found by solving a 4$^{th}$ degree equation, namely Eq.~\eqref{eq:app:varphi_OPT_4th_degree_generic}. However, for many input states, a simple analytical solution is possible, as detailed in Appendix \ref{subsec:app:Opt_varphi_single}. For the input state considered here and by employing Eq.~\eqref{eq:varphi_opt_coh_plus_sqz_vac_sing_det} we find the working points $\varphi_{opt} \approx0.87\pi$ and $\varphi_{opt}\approx1.12\pi$ (see Fig.~\ref{fig:Phase_sens_SING_3D_plot}).


Having access to an external phase reference and assuming a single internal phase shift \eqref{eq:one_phase_shift_CONVENTION}, not only changes the best achievable performance to $\Delta\varphi_{QCRB}^{(i)}$ given in Eq.~\eqref{eq:Delta_varphi_QCRB_i_DEFINTION}, but usually also favors unbalanced scenarios \cite{Ata20,Ata22}. When it comes to optimizing the phase sensitivity of a MZI using a BHD scheme (see Section \ref{subsec:BHD_scheme}), we have to optimize a function depending on the local oscillator phase ($\phi_L$), too \emph{i. e.}  $\Delta\varphi_{hom}({\vartheta'},\varphi,\phi_L)$. However, quite often the local oscillator phase is obvious (for example in phase with the input coherent source, by setting $\phi_L=\theta_\alpha$). We thus assume $\phi_L$ already at its optimum value and proceed as before to minimize the function $\Delta\varphi_{hom}({\vartheta'},\varphi)$.

\begin{figure}
\includegraphics[width=0.49\textwidth]{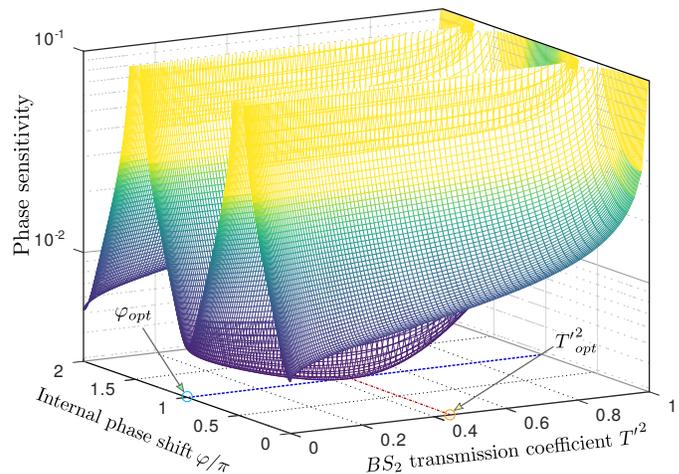}
\caption{Example plot of the phase sensitivity $\Delta\varphi_{hom}$ given in Eq.~\eqref{eq:Delta_varphi_hom_1phase_generic} for a BHD scheme versus the total internal phase shift $\varphi$ and the $BS_2$ transmission coefficient, ${T'}^2$. We consider the coherent plus squeezed vacuum input state given in Eq.~\eqref{eq:psi_in_coh_plus_sqz_vac} and $BS_1$ is optimized via QFI yielding $T\approx\sqrt{0.55}$. The optimum phase sensitivity is found for the working point(s) $\varphi_{opt}=\pi+2k\pi$ ($k\in\mathbb{Z}$) and for the $BS_2$ transmission coefficient $T'_{opt}\approx\sqrt{0.45}$. Parameters used: $\vert\alpha\vert=10^2$, $r=1.2$, and $2\theta_\alpha-\theta=0$. }
\label{fig:Phase_sens_HOM_3D_plot}
\end{figure}

In Fig.~\ref{fig:Phase_sens_HOM_3D_plot} give an example of phase sensitivity when an external phase reference is available. We consider the same input state given in Eq.~ \eqref{eq:psi_in_coh_plus_sqz_vac} and a BHD scheme. From the single-parameter QFI we can deduce the optimal transmission coefficient of the first BS, namely $T\approx\sqrt{0.55}$ \cite{Ata22}. By optimizing the second beam splitter we find, as expected an unbalanced beam splitter featuring $T'_{opt}\approx\sqrt{0.45}$ for $BS_2$. Details on this type of optimization are found in Appendix \ref{subsec:app:Opt_Tprime_homodyne}.

As for the working point for a MZI employing a BHD scheme and a single internal phase shift, quite often (see details in Appendix \ref{subsec:app:Opt_varphi_homodyne}) the working point is found to be $\varphi_{opt}=\pi+2k\pi$ ($k\in\mathbb{Z}$).

\section{Optimized phase sensitivity performance with some input states}
\label{sec:detector_performance}
In this section, we assess the performance of the three considered detection schemes for a number of relevant input states.

We start our discussion with Gaussian states. The ones already considered in the literature are discussed briefly while the more complicated case of the squeezed-coherent plus squeezed-coherent input state is detailed at length. We then go on to discuss an interesting non-Gaussian input, namely the coherent plus Fock state.

\subsection{Coherent plus squeezed vacuum input}
\label{subsec:opt_perf_coh_sqzvac}
One of the most widely used states both in theoretical and experimental quantum-enhanced metrology is the coherent plus squeezed vacuum  input \cite{Cav81,Dem15,API18,Pez08,Xia87},
\begin{equation}
\label{eq:psi_in_coh_plus_sqz_vac}
\vert\psi_{in}\rangle=\vert\alpha_1\xi_0\rangle,
\end{equation}
where the coherent (or Glauber) state in port $1$ is obtained by applying the displacement or Glauber operator \cite{GerryKnight,MandelWolf},
\begin{equation}
\label{eq:displacement_operator_def}
\hat{D}_1\left(\alpha\right)=e^{\alpha\hat{a}_1^\dagger-\alpha^*\hat{a}_1}
\end{equation}
\emph{i. e. } $\ket{\alpha_1}=\hat{D}_1\left(\alpha\right)\ket{0_1}$ with $\alpha=\vert\alpha\vert e^{i\theta_\alpha}$. The squeezed vacuum in port $0$, \emph{i. e.} ${\ket{\xi_0}=\hat{S}_0\left(\xi\right)\vert0_0\rangle}$, is obtained by applying the squeezing operator \cite{GerryKnight,Yue76}
\begin{equation}
\label{eq:Squeezing_operator}
\hat{S}_0\left(\xi\right)=e^{\frac{1}{2}\left(\xi^*\hat{a}_0^2-\xi(\hat{a}_0^\dagger)^2\right)},
\end{equation}
where $\xi=re^{i\theta}$. Usually $r\in\mathbb{R}^{+}$ is called the squeezing factor and ${\theta}$ denotes the phase of the squeezed state.

The optimum performance of this input state is obtained by imposing the input PMC \cite{Ata19,Gar17,API18}
\begin{equation}
\label{eq:PMC_coh_plus_sqz-vac}
2\theta_\alpha-\theta=0.
\end{equation}
When no external phase reference is available, the relevant QFI is the two-parameter one and it reaches its maximum when $BS_1$ is  balanced (\emph{i. e.} $T_{opt}=1/\sqrt{2}$) yielding $\mathcal{F}^{(2p)}_{max}=\vert\alpha\vert^2+\sinh^2r$ \cite{Pez08,API18,Dem15}. As discussed in previous works \cite{API18,Ata19,Gar17}, the two considered detection schemes falling into this scenario (Sec.~\ref{subsec:diff_intensity_mode} and \ref{subsec:single_mode}) are slightly suboptimal in respect with the QCRB from Eq.~\eqref{eq:Delta_varphi_QCRB_2p_DEFINTION} and their optimum performance is found for a balanced MZI \cite{Ata20} (see also Figs.~\ref{fig:Phase_sens_3D_difference}-\ref{fig:Phase_sens_SING_3D_plot}).

\begin{figure}
	\includegraphics[width=0.49\textwidth]{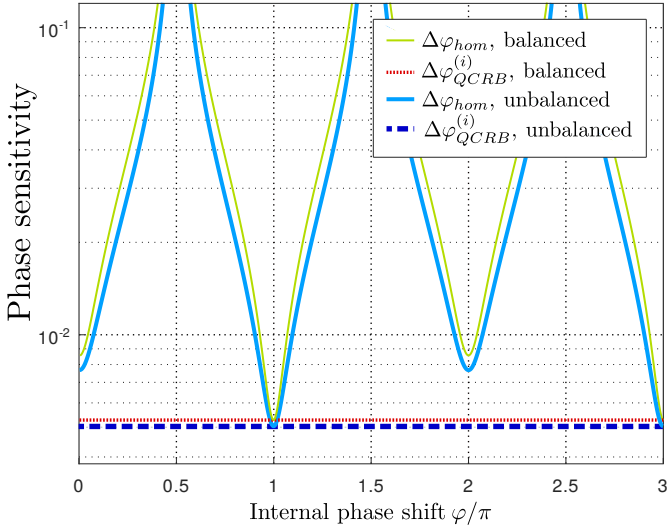}
	\caption{Phase sensitivity for a squeezed-coherent plus squeezed-vacuum input state in the balanced and unbalanced  scenarios with a BHD scheme. An unbalanced MZI is able to slightly outperform its balanced counterpart. Parameters used: $|\alpha|=50$, $r=1.2$ and $z=0.6$. The PMC employed is the optimum one given in Eq.~\eqref{eq:PMC_sqz-coh_plus_sqz-vac}.}
	\label{fig:Phase_sens_sqzcoh_sqzvac_alpha50_r1_2_z06-08}
\end{figure}

When an external phase reference is available, the relevant QFI is the single-parameter one,  $\mathcal{F}^{(i)}$, and its maximum value can be computed in closed form (see eq.~(F4) in reference \cite{Ata20}). This maximum is usually found for $BS_1$ unbalanced \cite{Ata20,Ata22}. The phase sensitivity optimization often also results in an unbalanced $BS_2$, as depicted in Fig.~\ref{fig:Phase_sens_HOM_3D_plot}.

It has been previously shown that, by employing a BHD scheme (see Sec.~\ref{subsec:BHD_scheme}) and an unbalanced MZI, one is able to approach the $\mathcal{F}^{(i)}$-induced QCRB, \emph{i. e.} $\Delta\varphi^{(i)}_{QCRB}$ (see Figs.~11 and 12 from reference \cite{Ata20}). 

Thus, only the availability of an external phase reference justifies the use of an unbalanced MZI in the case of the input state \eqref{eq:psi_in_coh_plus_sqz_vac}.

\subsection{Squeezed-coherent plus squeezed vacuum input}
\label{subsec:opt_perf_sqzcoh_sqzvac}
Consider now the squeezed-coherent plus squeezed vacuum input state \cite{Par95,Pre19,Ata20},
\begin{equation}
\label{eq:psi_in_sqzcoh_plus_sqz_vac}
\vert\psi_{in}\rangle=\vert(\alpha\zeta)_1\xi_0\rangle
=\hat{D}_1\left(\alpha\right)\hat{S}_1\left(\zeta\right)\hat{S}_0\left(\xi\right)\vert0\rangle {\red ,}
\end{equation}
and the squeezer in input port $1$ is characterized by ${\zeta=ze^{i\phi}}$. 
%
\begin{figure}
	\includegraphics[width=0.49\textwidth]{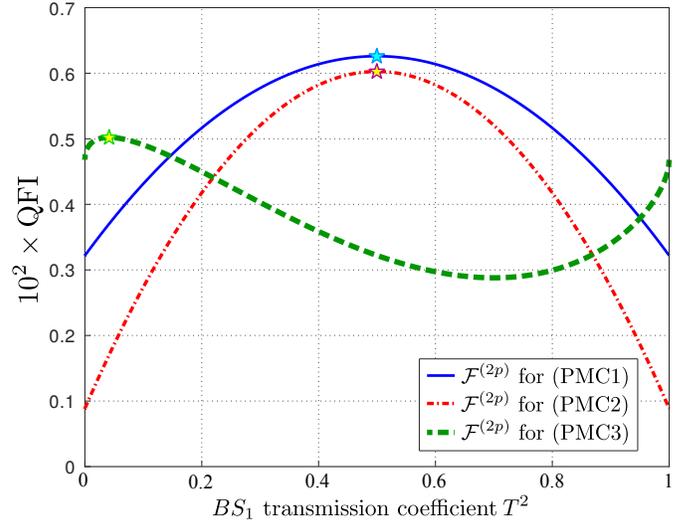}
	\caption{The two-parameter QFI ($\mathcal{F}^{(2p)}$) versus the transmission coefficient of the first beam splitter ($T^2$). We consider the input state given in Eq.~\eqref{eq:psi_in_sqzcoh_plus_sqzcoh} and compare the three input PMCs in the low-coherent regime. Parameters used: $ |\alpha|=2.2$, $|\beta|=1.4$, $r=1.2$, $z=0.6$. Stars mark the maximum of each QFI curve.}
	\label{fig:QFI_sqzcoh_sqzcoh_alpha2_2_beta1_4_PMC123}
\end{figure}
%
All QFIs are maximized if we impose the input PMC \cite{Ata20},
\begin{equation}
\label{eq:PMC_sqz-coh_plus_sqz-vac}
\left\{
\begin{array}{l}
2\theta_\alpha-\theta=0\\
2\theta_\alpha-\phi=\pm\pi.
\end{array}
\right.
\end{equation}
When no external phase reference is available, the optimum two-parameter QFI is found in the balanced case \cite{Ata20} yielding
 $\mathcal{F}^{(2p)}_{max}=\vert\alpha\vert^2+\sinh^2(r+z)$ \cite{Ata19}. For single- and difference-intensity detection schemes the optimum phase sensitivity is indeed found for a balanced MZI \cite{Ata20}.

Similar to the discussion from Section \ref{subsec:opt_perf_coh_sqzvac}, the availability of an external phase reference suggests a single-parameter QFI, $\mathcal{F}^{(i)}$, that is maximized in a non-balanced scenario \cite{Ata20,Ata22}. 

In Fig.~\ref{fig:Phase_sens_sqzcoh_sqzvac_alpha50_r1_2_z06-08}, we plot the phase sensitivity for the input state \eqref{eq:psi_in_sqzcoh_plus_sqz_vac} with a BHD scheme. After optimizing the non-balanced case we get $T_{opt}\approx\sqrt{0.71}\approx0.84$ and $T'_{opt}\approx\sqrt{0.28}\approx0.53$. As seen from Fig.~\ref{fig:Phase_sens_sqzcoh_sqzvac_alpha50_r1_2_z06-08}, the unbalanced scenario slightly outperforms the balanced one and we gain approximatively $4$ \% at the peak phase sensitivity.

We conclude the for the input state \eqref{eq:psi_in_sqzcoh_plus_sqz_vac}, similar to the coherent plus squeezed vacuum input, only the availability of an external phase reference justifies the phase sensitivity optimization via an unbalanced MZI.

\begin{figure}
	\includegraphics[width=0.49\textwidth]{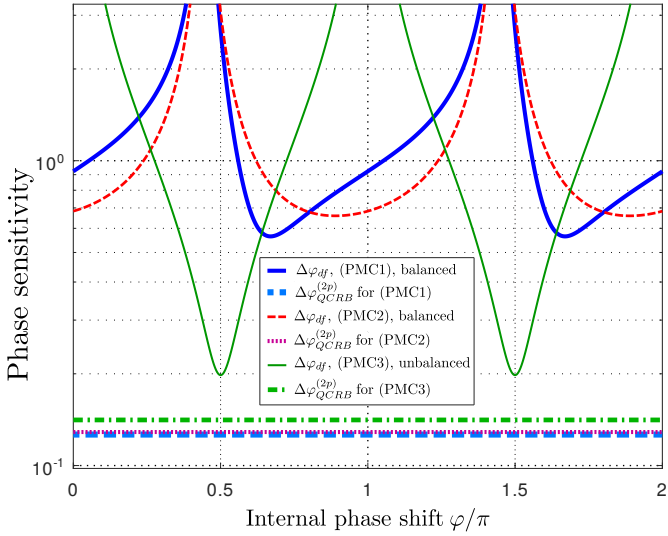}
	\caption{Difference-intensity detection phase sensitivity in the low-coherent intensity regime for a squeezed-coherent plus squeezed coherent input state \eqref{eq:psi_in_sqzcoh_plus_sqzcoh}. Parameters used: $\vert\alpha\vert=2.2$, $\vert\beta\vert=1.4$, $r=1.2$ and $z=0.6$. }
	\label{fig:Phase_sens_sqzcoh_sqzcoh_2_2_beta_1_4_diff_PMC123}
\end{figure}

\subsection{Squeezed-coherent plus squeezed-coherent input}
\label{subsec:opt_perf_sqzcoh_sqzcoh}
We consider now the squeezed-coherent plus squeezed-coherent input state \cite{Spa15,Ata19,Ata22},
\begin{equation}
\label{eq:psi_in_sqzcoh_plus_sqzcoh}
\vert\psi_{in}\rangle=\vert(\alpha\zeta)_1(\beta\xi)_0\rangle,
\end{equation}
where for port $0$ we have $\ket{(\beta\xi)_0}=\hat{D}_0\left(\beta\right)\hat{S}_0\left(\xi\right)\vert0\rangle$ and $\beta=\vert\beta\vert e^{i\theta_\beta}$. Optimum performance of this input state in terms of QFI imposes one of the three input phase matching conditions, namely \cite{Ata19,Ata22}:
\begin{equation}
\label{eq:PMC1_optimal_sqzcoh_sqzcoh}
\text{(PMC1)}
\left\{
\begin{array}{l}
2\theta_\alpha-\theta=0\\
\phi-\theta=\pm\pi\\
\theta_\alpha-\theta_\beta=0,
\end{array}
\right.
\end{equation}
\begin{equation}
\label{eq:PMC2_optimal_sqzcoh_sqzcoh}
\text{(PMC2)}
\left\{
\begin{array}{l}
2\theta_\alpha-\theta=0\\
\phi-\theta=0\\
\theta_\alpha-\theta_\beta=0,
\end{array}
\right.
\end{equation}
or
\begin{equation}
\label{eq:PMC3_optimal_sqzcoh_sqzcoh}
\text{(PMC3)}
\left\{
\begin{array}{l}
2\theta_\alpha-\theta=0\\
\phi-\theta=\pm\pi\\
\theta_\alpha-\theta_\beta=\frac{\pi}{2}.
\end{array}
\right.
\end{equation}

\begin{figure}
	\includegraphics[width=0.49\textwidth]{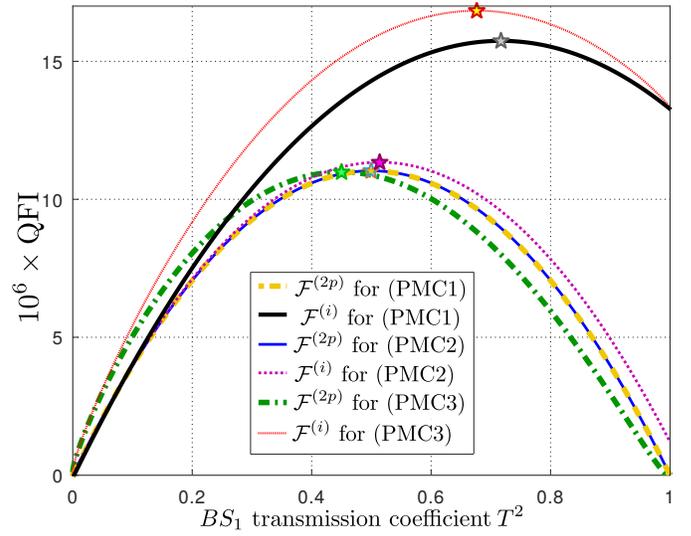}
	\caption{The single- and two-parameter QFI versus the transmission coefficient of the first beam splitter for the three input PMCs in the high-intensity coherent regime. A squeezed-coherent plus squeezed-coherent input state \eqref{eq:psi_in_sqzcoh_plus_sqzcoh} is considered. Parameters used: $\vert\alpha\vert=10^3$, $\vert\beta\vert=50$, $r=1.2$, $z=0.6$. Stars mark the maximum of each QFI curve.}
	\label{fig:QFI_sqzcoh_sqzcoh_alpha1000_beta50_PMC123}
\end{figure}

\noindent We start our discussion in the low-coherent intensity regime, \emph{i. e.} when 
\begin{equation}
\label{eq:sqzcoh_sqzcoh_LOW_coherent}
\{\vert\alpha\vert^2,\vert\beta\vert^2\}\approx\{\sinh^2r,\sinh^2z\}.
\end{equation}
For the difference-intensity detection scheme, the relevant QFI is $\mathcal{F}^{(2p)}$. In Fig.~\ref{fig:QFI_sqzcoh_sqzcoh_alpha2_2_beta1_4_PMC123} we plot the aforementioned QFI for the three considered PMCs versus the transmission coefficient of the first beam splitter. The optimum performance is thus expected by employing a balanced MZI and (PMC1). The second best performance is expected if we employ (PMC2) still in the balanced case, while the worst performance is predicted for (PMC3) and a heavily unbalanced MZI ($T^{(2p)}_{opt}\approx\sqrt{0.04}=0.2$).

In Fig.~\ref{fig:Phase_sens_sqzcoh_sqzcoh_2_2_beta_1_4_diff_PMC123}, we plot the actual performance in terms of phase sensitivity for a difference-intensity detection scheme in the low-intensity coherent regime. Although the best performance (based on the QFI-induced QCRB) is expected for (PMC1), this is not what actually happens. The input (PMC3) is found to yield the best performance (solid thin green curve) and the optimum transmission coefficient for $BS_2$ is found to be $T'_{opt}\approx\sqrt{0.498}=0.706$. The \emph{actual} performance of a balanced MZI with either (PMC1) or (PMC2) is found to be inferior to the unbalanced MZI scenario.

While the low-intensity regime is interesting for a number of applications involving usually delicate or light-affected samples (live biological cells, retina samples etc.), the high intensity regime 
\begin{equation}
\label{eq:sqzcoh_sqzcoh_HIGH_coherent}
\{\vert\alpha\vert^2,\vert\beta\vert^2\}\gg\{\sinh^2r, \sinh^2z\}
\end{equation}
is also interesting especially for high-precision measurements. In all our calculations, besides the constraint \eqref{eq:sqzcoh_sqzcoh_HIGH_coherent} we will also impose the second coherent source to be weak in respect with the first one (\emph{i. e.} $\vert\alpha\vert^2\gg\vert\beta\vert^2\gg\{\sinh^2r, \sinh^2z\}$). This constraint is set to be more realistic, since experiments usually have one high-power laser. While the low-intensity regime favours the difference-intensity detection scheme, in the high intensity coherent regime, the single-mode intensity and BHD schemes are experimentally preferred \cite{Gar17,Ata19}.

\begin{figure}
	\includegraphics[width=0.49\textwidth]{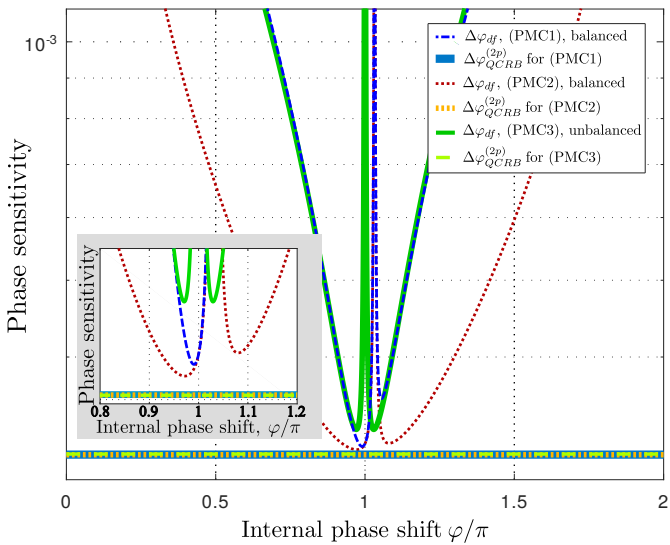}
	\caption{Phase sensitivity for the squeezed-coherent plus squeezed coherent input in the high-coherent regime using a single-mode intensity detection. All three input PMCs yield suboptimal performance, with (PMC2) yielding the best result ($\Delta\varphi_{sg}^{opt}=3.084\times10^{-4}$ at $\varphi_{opt}=0.99\pi$) while (PMC3) the worst ($\Delta\varphi_{sg}^{opt}=3.241\times10^{-4}$ at $\varphi_{opt}=0.97\pi$). Inset: zoom around $\varphi=\pi$, where all three phase sensitivities peak. Parameters used: $\vert\alpha\vert=10^3$, $\vert\beta\vert=50$, $r=1.2$, $z=0.6$.}
	\label{fig:Phase_sens_sqzcoh_sqzcoh_1000PMC123_beta50_sg}
\end{figure}

In Fig.~\ref{fig:QFI_sqzcoh_sqzcoh_alpha1000_beta50_PMC123} we plot both the single- and two-parameter QFI versus the transmission coefficient of the first beam splitter for the three input PMCs in the high-intensity coherent regime \cite{Ata22}.

For the single-mode intensity detection the relevant QFI is still the two-parameter one, $\mathcal{F}^{(2p)}$. It becomes obvious from Fig.~\ref{fig:QFI_sqzcoh_sqzcoh_alpha1000_beta50_PMC123} that the optimum $\mathcal{F}^{(2p)}_{max}$ is roughly identical for all three input PMCs. (One finds $\mathcal{F}^{(2p)}_{max}=11.024\times10^6$ for (PMC1), $\mathcal{F}^{(2p)}_{max}=11.031\times10^6$ for (PMC2) while for (PMC3) we have $\mathcal{F}^{(2p)}_{max}=10.987\times10^6$.) However, while for (PMC1) and (PMC2) the optimum is found in the balanced case, for (PMC3) it is found in an unbalanced scenario  with $T_{opt}=\sqrt{0.45}$.

In Fig.~\ref{fig:Phase_sens_sqzcoh_sqzcoh_1000PMC123_beta50_sg} we depict the performance of a single-mode intensity detection scheme for a squeezed-coherent plus squeezed-coherent input state for all three PMCs. While all PMCs yield slightly sub-optimal and quite equivalent performance, (PMC3) seems to perform worst. 

\begin{figure}
	\includegraphics[width=0.49\textwidth]{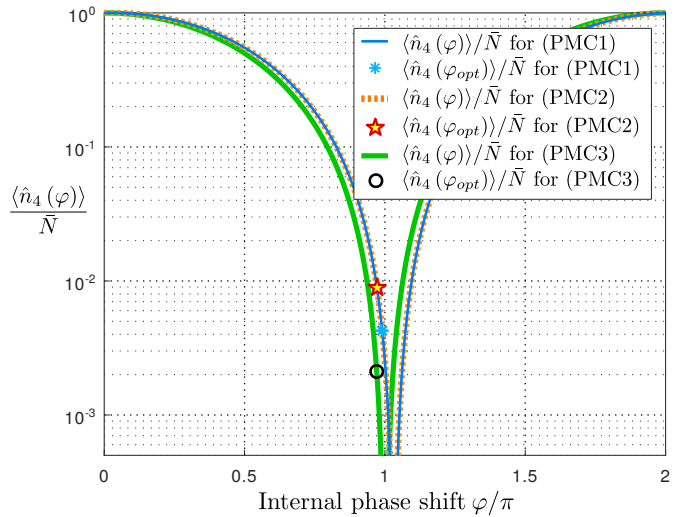}
	\caption{Extinction rates $\braket{\hat{n}_4\left(\varphi\right)}/\bar{N}$ for the squeezed-coherent plus squeezed coherent input in the high-coherent regime using a single-mode intensity detection. For (PMC3) and an unbalanced MZI we find $\braket{\hat{n}_4\left(\varphi_{sg}^{opt}\right)}/\bar{N}=2.1128\times10^{-3}$; for a balanced MZI and (PMC1) we find $\braket{\hat{n}_4\left(\varphi_{sg}^{opt}\right)}/\bar{N}=4.257\times10^{-3}$ while for (PMC2) we have $\braket{\hat{n}_4\left(\varphi_{sg}^{opt}\right)}/\bar{N}=8.876\times10^{-3}$.}
	\label{fig:n4_extinction_sqzcoh_sqzcoh_alpha1000_beta50}
\end{figure}

However, this slight disadvantage can be compensated by the advantage gained in the extinction rate $\braket{\hat{n}_4\left(\varphi\right)}/\bar{N}$ at the optimum working point \emph{i. e.} $\braket{\hat{n}_4\left(\varphi_{opt}\right)}/\bar{N}$. Indeed, especially in the high intensity regime, it is desirable to have the output port nearly ``dark''. In Fig.~\ref{fig:n4_extinction_sqzcoh_sqzcoh_alpha1000_beta50} we plot the extinction rates for all three PMCs as well as the extinction rates at their respective optimum working points. We conclude that although (PMC3) and an unbalanced MZI did not yield the best phase sensitivity (as depicted in Fig.~\ref{fig:Phase_sens_sqzcoh_sqzcoh_1000PMC123_beta50_sg}), it outperformed the other input PMCs using a balanced MZI in terms of extinction rate at the optimum working point.

\begin{figure}
	\includegraphics[width=0.49\textwidth]{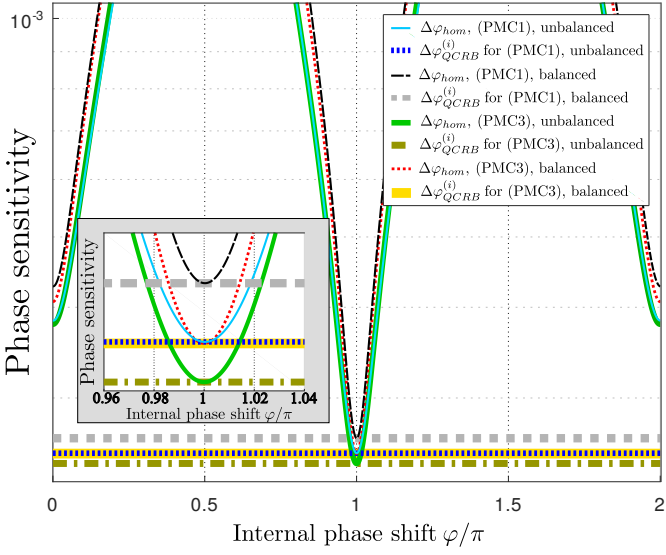}
	\caption{Phase sensitivity for the squeezed-coherent plus squeezed coherent input in the high-coherent regime. Unbalancing the MZI brings an advantage in terms of best phase sensitivity both in the case of (PMC3) (thick solid green curve versus thin dotted red curve) as well as (PMC1) (thin solid light-blue curve versus thin dashed black curve). Parameters used: $\vert\alpha\vert=10^3$, $\vert\beta\vert=50$, $r=1.2$, and $z=0.6$. Inset: zoom around the $\varphi=\pi$ working point.}
	\label{fig:Phase_sens_sqzcoh_sqzcoh_1000PMC13_beta50_hom}
\end{figure}

We set our attention now to the case when the detection scheme has access to an external phase reference. As mentioned previously, the relevant QFI in this scenario is the single-parameter one, $\mathcal{F}^{(i)}$ \cite{Jar12,Ata20}.

From the QFI prediction (see Fig.~\ref{fig:QFI_sqzcoh_sqzcoh_alpha1000_beta50_PMC123}), the best performance for the input state \eqref{eq:psi_in_sqzcoh_plus_sqzcoh} is expected if we employ (PMC3) and an unbalanced MZI with $BS_1$ featuring $T_{opt}=\sqrt{0.67}$. The next best performance is expected for (PMC1) in which case $T_{opt}=\sqrt{0.7}$. We set to verify these predictions in Fig.~\ref{fig:Phase_sens_sqzcoh_sqzcoh_1000PMC13_beta50_hom}. Indeed, employing (PMC3) and an unbalanced MZI (the optimal $BS_2$ transmission coefficient is found to be $T'_{opt}=\sqrt{0.278}$) we get the optimum phase sensitivity $\Delta\varphi_{hom}$ (thick solid green curve) reaching its optimal value $\Delta\varphi_{hom}^{opt}=2.437\times10^{-4}$ at the working point $\varphi_{opt}=\pi$. Imposing a balanced MZI for the same input PMC (thin dotted red curve) degrades the optimum phase sensitivity and the best value at the working point is found to be $\Delta\varphi_{hom}^{opt}=2.515\times10^{-4}$.

As predicted by the QFI plot (see Fig.~\ref{fig:QFI_sqzcoh_sqzcoh_alpha1000_beta50_PMC123}), the next best performance is potentially given by the input state \eqref{eq:psi_in_sqzcoh_plus_sqzcoh} and (PMC1). Indeed, one finds $\Delta\varphi_{hom}$ (thin solid light-blue curve) reaching the optimum phase sensitivity $\Delta\varphi_{hom}^{opt}= 2.517\times10^{-4}$  at the optimum working point $\varphi_{opt}=\pi$. Imposing a balanced MZI for the same input PMC (thin dashed black curve) degrades the optimum phase sensitivity to $\Delta\varphi_{hom}^{opt}=2.64\times10^{-4}$.

We also depicted in Fig.~\ref{fig:Phase_sens_sqzcoh_sqzcoh_1000PMC13_beta50_hom} all corresponding QCRBs, in order to assess the sub-optimality of each considered scenario. From the inset one can see that all scenarios are nearly optimal, the thickness of the lines not allowing the visibility of the minute sub-optimality of each scheme. For example the best performance, as stated previously belongs to a non-balanced MZI with (PMC3) and we found $\Delta\varphi_{hom}^{opt}=2.43732\times10^{-4}$. The corresponding QCRB is $\Delta\varphi_{QCRB}^{(i)}=2.43729\times10^{-4}$.

In ref.~\cite{Spa15} it was claimed that for a squeezed-coherent plus squeezed-coherent input state ``Unbalanced devices may be also considered, which however lead to inferior performances''. As discussed in this section, at least for some input parameters, this claim cannot be sustained.

We conclude that when it comes the squeezed-coherent plus squeezed-coherent input state given by Eq.~\eqref{eq:psi_in_sqzcoh_plus_sqzcoh}, it is more difficult to point out when an unbalanced MZI is able to outperform its balanced counterpart. Indeed, while in the previous sections (Sec.~\ref{subsec:opt_perf_coh_sqzvac} and \ref{subsec:opt_perf_sqzcoh_sqzvac}) the availability of an external phase reference justified unbalancing the MZI, for the input state \eqref{eq:psi_in_sqzcoh_plus_sqzcoh} even not having access to an external phase reference might justify this choice.

\subsection{Coherent plus Fock input}
\label{subsec:opt_perf_coh_Fock}
As a last example we consider the coherent plus Fock input state \cite{Birrittella2012,Birrittella2021},
\begin{equation}
\label{eq:psi_in_Fock_plus_coherent}
\vert\psi_{in}\rangle=\vert\alpha_1n_0\rangle,
\end{equation}
where $\hat{n}_0\ket{n_0}=n\ket{n_0}$. 

If no external phase reference is available, the optimum in terms of phase sensitivity is found in the balanced case \cite{Ata22}. The optimization of the second BS yields a similar result. However, if an external phase reference is available, from the single-parameter QFI \eqref{eq:Delta_varphi_QCRB_i_DEFINTION} we obtain the optimum transmission coefficient for $BS_1$, namely \cite{Ata22},
\begin{equation}
\label{eq:T_i_opt_C2prime_C4prime_ZERO_Fock_plus_coherent}
T^{(i)}_{opt}
=\sqrt{\frac{1}{2}+\frac{\vert\alpha\vert^2}{2n\left(1+2\vert\alpha\vert^2\right)}}.
\end{equation}
The optimum working point is found, as expected from Eq.~\eqref{eq:app:varphi_OPT_pi_generic}, among multiples of $\pi$, irrespective on the values of $n$ and $\alpha$ (assuming $\alpha\neq0$).

\begin{figure}
	\includegraphics[width=0.49\textwidth]{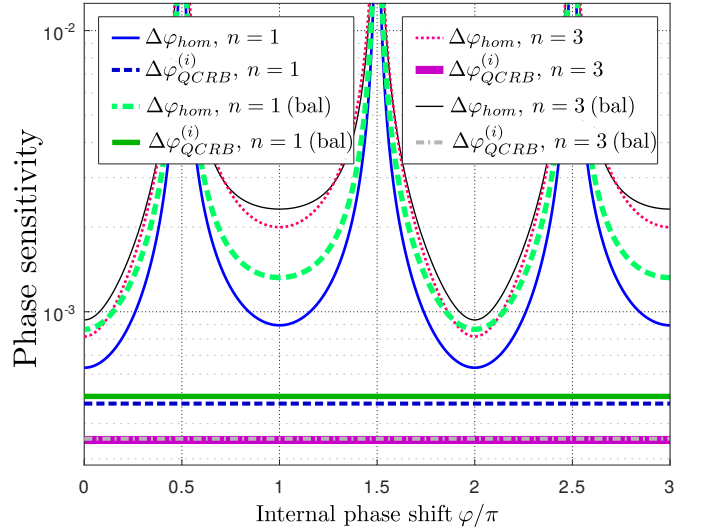}
	\caption{Phase sensitivity for a coherent plus Fock input state. The unbalanced cases outperform the balanced ones (labelled ``bal'' in the legend) for both $n=1$ and $n=3$. Parameter used: $\vert\alpha\vert=10^3$. For the unbalanced scenarios, the first BS is optimized following Eq.~\eqref{eq:T_i_opt_C2prime_C4prime_ZERO_Fock_plus_coherent}, while the second one is optimized via Eq.~\eqref{eq:tan_vartheta_prime_OPT_non-entangled_at_varphi_OPT}.}
	\label{fig:Phase_sens_coh_Fock_100_n1}
\end{figure}

In Fig.~\ref{fig:Phase_sens_coh_Fock_100_n1} we plot the phase sensitivity for a coherent plus Fock input and a BHD scheme, both in the balanced and unbalanced scenarios. In order to be realistic, we employ the high coherent regime, $\vert\alpha\vert^2\gg n$.

For $\vert\alpha\vert=10^3$ and $n=1$ we have from Eq.~\eqref{eq:T_i_opt_C2prime_C4prime_ZERO_Fock_plus_coherent} $T_{opt}\approx\sqrt{0.75}=0.866$. Optimizing the second beam splitter via Eq.~\eqref{eq:tan_vartheta_prime_OPT_non-entangled_at_varphi_OPT} yields a transmission coefficient $T'_{opt}\approx\sqrt{0.107}=0.328$. As expected, the unbalanced MZI scenario (solid blue curve in  Fig.~\ref{fig:Phase_sens_coh_Fock_100_n1}, lowest among the wavy lines) outperforms in terms of phase sensitivity the balanced one (thick dashed light-green curve). Both scenarios, though, are suboptimal in respect with the QCRBs (thick dashed dark-blue and, respectively, thick solid dark green horizontal lines).

Keeping the coherent amplitude fixed but increasing the photon number for the Fock state to $n=3$ results in a less important advantage for the unbalanced (dotted pink-red line) versus the balanced MZI (thin black line) in terms of phase sensitivity performance. This result is not surprising, since $T^{(i)}_{opt}$ from Eq.~\eqref{eq:T_i_opt_C2prime_C4prime_ZERO_Fock_plus_coherent} implies that $BS_1$ becomes balanced as $n$ grows indefinitely. While the unbalanced scenario for $n=3$ (we find $T_{opt}\approx\sqrt{0.577}\approx0.76$ and $T'_{opt}\approx\sqrt{0.3}\approx0.55$) still outperforms the balanced one, the sub-optimality of both scenarios increases in respect with the case $n=1$. This becomes obvious when comparing the performances for $n=3$ with the two corresponding QCRBs (thick solid violet and, respectively, dash-dotted gray lines).  We discuss more about this sub-optimality in Sec.~\ref{subsec:opt_perf_coh_sqzvac_VERSUS_coh_Fock}.

We conclude that for the coherent plus Fock input state \eqref{eq:psi_in_Fock_plus_coherent}, only the availability of an external phase reference can justify unbalancing the MZI.

\subsection{Performance comparison for two input states: coherent plus squeezed vacuum versus coherent plus Fock}
\label{subsec:opt_perf_coh_sqzvac_VERSUS_coh_Fock}

In reference \cite{Ata20} it was shown that a unbalanced MZI can outperform a balanced one in terms of phase sensitivity if a coherent plus squeezed vacuum is applied at its input. In Fig.~\ref{fig:Phase_sens_coh_Fock_100_n1} we showed that an unbalanced MZI can show an advantage in terms of phase sensitivity over a balanced one in terms of phase sensitivity for a coherent plus Fock input state.

In reference \cite{Ata22}, the two input states \emph{i. e.} Eq.~\eqref{eq:psi_in_coh_plus_sqz_vac} and Eq.~\eqref{eq:psi_in_Fock_plus_coherent} were compared in terms of single- and two-parameter QFI (see Fig.~12 from the aforementioned reference). It was found that their performance has a similar scaling in respect with the input resources, with a slight advantage for the coherent plus squeezed vacuum input state. From the QFI performance only, one cannot decide which of the states Eq.~\eqref{eq:psi_in_coh_plus_sqz_vac} and Eq.~\eqref{eq:psi_in_Fock_plus_coherent} shows a practical advantage. We will assess this question in the following.

\begin{figure}
	\includegraphics[width=0.49\textwidth]{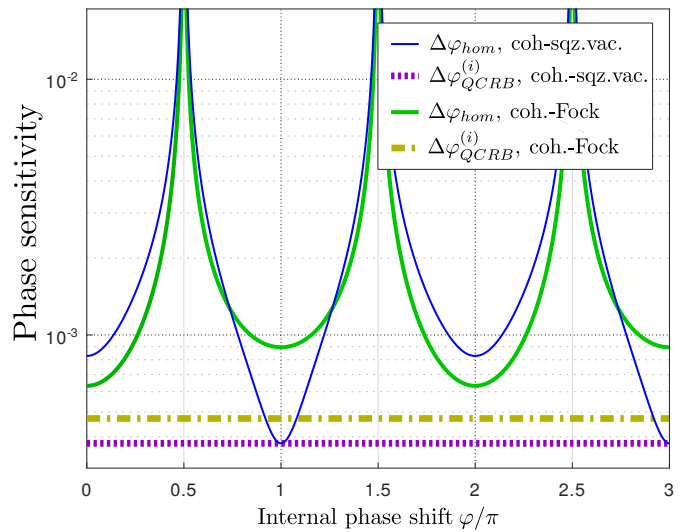}
	\caption{Phase sensitivity comparison between coherent plus squeezed vacuum \eqref{eq:psi_in_coh_plus_sqz_vac} and coherent plus Fock \eqref{eq:psi_in_Fock_plus_coherent} input states for an unbalanced MZI and a BHD scheme. While the phase sensitivity for coherent plus squeezed vacuum input is almost optimal, for a coherent plus Fock input state it remains largely suboptimal. Parameters used: $\vert\alpha\vert=10^3$. For the coherent plus Fock input we used $n=1$ while for the coherent plus squeezed vacuum input we imposed $r=0.882$ so that $n\approx\sinh^2r$.}
	\label{fig:Coh_sqzvac_vs_coh_Fock_alpha1k_n_1}
\end{figure}

For both difference-intensity and single mode detection schemes we found that the coherent plus Fock input state yielded largely sub-optimal results, while the coherent plus squeezed vacuum, as already reported in the literature, yielded nearly optimal results \cite{Ata19,Gar17,Dem15,Ata20}.

In Fig.~\ref{fig:Coh_sqzvac_vs_coh_Fock_alpha1k_n_1} we plot the two aforementioned states for unbalanced MZIs and a BHD scheme. For each input state, the MZI is optimized following the already described procedure. We keep the same parameters employed in reference \cite{Ata22}. It ought to be mentioned that Fig.~\ref{fig:Coh_sqzvac_vs_coh_Fock_alpha1k_n_1} depicts the most favourable scenario for the coherent plus Fock input, namely for $n=1$. For $n=3$, this input state becomes largely suboptimal (as depicted in Fig.~\ref{fig:Coh_sqzvac_vs_coh_Fock_alpha1k_n_1}), while the coherent plus squeezed vacuum remains almost optimal, even for increasing squeezeing factors (see \emph{e. g.} Fig.~12 from reference \cite{Ata20}).

We conclude that while in terms of the QFI-induced QCRB the two states Eq.~\eqref{eq:psi_in_coh_plus_sqz_vac} and Eq.~\eqref{eq:psi_in_Fock_plus_coherent} show a roughly similar performance, in terms of realistic phase sensitivity the coherent plus squeezed vacuum outperforms the coherent plus Fock input for all three detection schemes considered in this work. This explains why practical implementations usually prefer the coherent plus squeezed vacuum state \cite{LIGO13,Tse2019}  when implementing sub-shot noise interferometry.

However, we would like to point out that Fock state interferometry has been shown to perform well with parity \cite{Birrittella2021,Campos2003} or double parity \cite{Zhong2021} detection. While for very low input average photon numbers this detection technique is feasible, for high-intensity input light, it becomes problematic.

\section{Conclusions}
\label{sec:conclusions}
In this paper we addressed the problem of phase sensitivity optimization for an unbalanced Mach-Zehnder interferometer with three different detection schemes.

We optimized the first beam splitter guided by the quantum Fisher information. We discussed both the single- and two-parameter QFI cases, thus being able to take into account the availability (or not) of an external phase reference. The optimum transmission coefficient of the second beam splitter was obtained as a result of a two variable function optimization problem, where the optimal working point of the interferometer had to be taken into account, too.

For detection schemes not having access to an external phase reference (single- or difference-mode detection schemes in our case) we found that the optimal phase sensitivity is almost always obtained by employing balanced $BS_2$. The optimum working point, though, was found to be detector-dependent.

When it comes to a detection scheme having access to an external phase reference, the balanced MZI scenario is rather the exception, not the rule. Unbalancing the MZI, almost always showed an advantage in terms of phase sensitivity.

Examples of non-balanced MZI optimizations were given for both Gaussian and non-Gaussian input states with a more in-depth discussion of the squeezed-coherent plus squeezed-coherent input state.

\begin{acknowledgments}
One of us (S. A.) would like to thank Wei Zhong for interesting discussions and suggestions for the improvement of this manuscript.

It is also acknowledged that this work has been supported by the Extreme Light Infrastructure Nuclear Physics (ELI-NP) Phase II, a project co-financed by the Romanian Government and the European Union through the European Regional Development Fund and the Competitiveness Operational Programme (1/07.07.2016, COP, ID 1334).

\end{acknowledgments}


\appendix

\begin{widetext}

\section{Single- and two-parameter QFI}
\label{sec:app:single_two_parameter_QFI}
When estimating a (single) parameter problem, the optimal solution is found by employing the quantum version of the Fisher information, \emph{i. e.} the quantum Fisher information (QFI) \cite{Helstrom1967,Helstrom1968,Holevo1973,Bra94}. For a thorough discussion of this topic, see reference \cite{Par09}.

When dealing with a multi-parameter estimation problem the single QFI approach has to be extended to a matrix form \cite{Lan13,Jar12,Ata20}. We thus replace the QFI by a $2\times2$ matrix: 
\begin{equation}
\label{eq:Fisher_matrix}
   \boldsymbol{\mathcal{F}}=
  \left[ {\begin{array}{cc}
  \mathcal{F}_{ss} & \mathcal{F}_{sd} \\
   \mathcal{F}_{ds} & \mathcal{F}_{dd} \\
  \end{array} } \right]
\end{equation}
having the following Fisher matrix elements \cite{Jar12,Lan13},
\begin{equation}
\label{eq:Fisher_matrix_elements}
\mathcal{F}_{ij}=4\Re\{\langle\partial_i\psi\vert\partial_j\psi\rangle-\langle\partial_i\psi\vert\psi\rangle
 \langle\psi\vert\partial_j\psi\rangle\}
\end{equation}
with $i, j\in \{s,d\}$ and we performed the variable changes $\varphi_s=\varphi_1+\varphi_2$ and $\varphi_d=\varphi_1-\varphi_2$. The quantum Cram\'er-Rao bound inequality implies \cite{Lan13,Pez15}
\begin{equation}
\label{eq:Sigma_matrix_greater_inv_Fisher_matrix}
  \left[ {\begin{array}{cc}
  (\Delta\varphi_s)^{2} & \text{Cov}(\varphi_s,\varphi_d) \\
   \text{Cov}(\varphi_s,\varphi_d) & (\Delta\varphi_d)^{2} \\
  \end{array} } \right]
  =\boldsymbol{\Sigma}\geq\boldsymbol{\mathcal{F}}^{-1}=\frac{1}{\mathcal{F}_{ss}\mathcal{F}_{dd}-\mathcal{F}_{sd}\mathcal{F}_{ds}}
  \left[ {\begin{array}{cc}
  \mathcal{F}_{dd} & -\mathcal{F}_{sd} \\
   -\mathcal{F}_{ds} & \mathcal{F}_{ss} \\
  \end{array} } \right].
\end{equation}
Generally, this matrix inequality \emph{i. e.} $\Sigma\geq\mathcal{F}^{-1}$ cannot be saturated for all components \cite{Par09}. However, we are solely interested in the difference-difference phase estimator, $\Delta\varphi_d$, thus the only inequality we are interested to saturate is
\begin{equation}
\label{eq:Delta_varphi_geq_Fisher_2p_dd}
(\Delta\varphi_d)^2\geq(\boldsymbol{\mathcal{F}}^{-1})_{dd}=\frac{\mathcal{F}_{ss}}{\mathcal{F}_{ss}\mathcal{F}_{dd}-\mathcal{F}_{sd}^2}
\end{equation}
and in order to simplify the writing we were led to introduce the definition from Eq.~\eqref{eq:Fisher_information_F_2p_DEFINITION}. We also used in the last equation the obvious fact that $\mathcal{F}_{ds}=\mathcal{F}_{sd}$. The single-parameter QFI, $\mathcal{F}^{(i)}$, can be computed from the Fisher matrix coefficients \cite{Ata20,Ata22},
\begin{equation}
\label{eq:F_i_is_Fss_plus_Fdd-2Fds}
\mathcal{F}^{(i)}=\mathcal{F}_{ss}+\mathcal{F}_{dd}-2\mathcal{F}_{sd}.
\end{equation}

\section{Difference-mode intensity detection calculations}
\label{sec:app:diff_calculations}
From the field operator transformations \eqref{eq:field_op_transf_a4_a5_vs_a0_a1} we have
\begin{equation}
\label{eq:Nd_fct_Nd_Zd_a0_a1}
\hat{N}_d=K_z\left({\hat{n}_0}-{\hat{n}_1}\right)
+K_x\left({\hat{a}_0}{\hat{a}_1^\dagger}+{\hat{a}_0^\dagger}{\hat{a}_1}\right)
+iK_y\left({\hat{a}_0}{\hat{a}_1^\dagger}-{\hat{a}_0^\dagger}{\hat{a}_1}\right)
\end{equation}
and employing the Schwinger pseudo-angular momentum operators we get the result from Eq.~\eqref{eq:Nd_fct_Nd_Re-Im_Zd_Jx_Jy_Jz}. Averaging the square of the operator $\hat{N}_d$ takes us to
\begin{eqnarray}
\label{eq:Average_Nd_2_fct_Kx_Ky_Kz_Jx_Jy_Jz}
\braket{\hat{N}_d^2}=
4K_x^2\braket{\hat{J}_x^2}+4K_y^2\braket{{\hat{J}_y^2}}+4K_z^2\braket{{\hat{J}_z^2}}
+4K_xK_y\left(\braket{\hat{J}_x{\hat{J}_y}}+\braket{{\hat{J}_y}\hat{J}_x}\right)
\nonumber\\
+4K_xK_z\left(\braket{\hat{J}_x{\hat{J}_z}}+\braket{{\hat{J}_z}\hat{J}_x}\right)
+4K_yK_z\left(\braket{{\hat{J}_y}{\hat{J}_z}}+\braket{{\hat{J}_z}{\hat{J}_y}}\right).
\end{eqnarray}
Using the above result as well as Eq.~\eqref{eq:Nd_fct_Nd_Re-Im_Zd_Jx_Jy_Jz} takes us to the variance given in Eq.~\eqref{eq:Nd_Variance_Kx_Ky_Kz_Jx_Jy_Jz}.

\section{Single-mode intensity detection calculations}
\label{sec:app:single_calculations}
From Eq.~\eqref{eq:field_op_transf_a4_a5_vs_a0_a1} we find the output number operator
\begin{equation}
\label{eq:a4_fct_Acoeffs_a0_a1}
{\hat{n}_4}
=|{\mathcal{A}}_{40}|^2{\hat{n}_0}
+|{\mathcal{A}}_{41}|^2{\hat{n}_1}
+{\mathcal{A}}^*_{40}{\mathcal{A}}_{41}{\hat{a}_0^\dagger}{\hat{a}_1}
+{\mathcal{A}}_{40}{\mathcal{A}}^*_{41}{\hat{a}_0}{\hat{a}_1^\dagger}
\end{equation}
and using relations \eqref{eq:Kd_Zd_constraints_A_coeffs} we end up with the result from Eq.~\eqref{eq:n4_fct_Nd_Re-Im_Zd_Jx_Jy_Jz}. The average of the number operator ${\hat{n}_4}$ is thus
\begin{equation}
\label{eq:a4_avg_fct_Kd_Zd_a0_a1_Re_compact}
\braket{\hat{n}_4}
=\frac{1+K_z}{2}\braket{\hat{n}_0}
+\frac{1-K_z}{2}\braket{\hat{n}_1}
+K_x\Re\left\{\braket{{\hat{a}_0}{\hat{a}_1^\dagger}}\right\}
-K_y\Im\left\{\braket{{\hat{a}_0}{\hat{a}_1^\dagger}}\right\}
\end{equation}
and by employing Eqs.~\eqref{eq:QO_BS_and_MZI:Schwinger_J1_operator}-\eqref{eq:N_is_a_dagger_a_0_plus_1} we end up with
\begin{equation}
\label{eq:n4_avg_fct_Nd_Re-Im_Zd_Jx_Jy_Jz}
\braket{\hat{n}_4}
=\frac{1}{2}\braket{\hat{N}}
+K_x\braket{\hat{J}_x}
+K_y\braket{\hat{J}_y}
+K_z{\braket{\hat{J}_z}}.
\end{equation}
Its derivative in respect with $\varphi$ is easily obtained and given in Eq.~\eqref{eq:del_n4_avg__del_varphi_fct_Jx_Jy_Jz}. In order to find ${\Delta^2\hat{n}_4}$, we first square the operator \eqref{eq:n4_fct_Nd_Re-Im_Zd_Jx_Jy_Jz} and using the fact that $[J_k,N]=0$ for $k=\{x,y,z\}$ we find
\begin{eqnarray}
{\hat{n}_4}^2
=\frac{1}{4}{\hat{N}^2}
+K_x^2{\hat{J}_x^2}
+K_y^2{\hat{J}_y^2}
+K_z^2{\hat{J}_z^2}
+K_x{\hat{J}_x}{\hat{N}}
+K_y{\hat{J}_y}{\hat{N}}
+K_z{\hat{J}_z}{\hat{N}}
\nonumber\\
+K_xK_y\left({\hat{J}_x}{\hat{J}_y}+{\hat{J}_y}{\hat{J}_x}\right)
+K_xK_z\left({\hat{J}_x}{\hat{J}_z}+{\hat{J}_z}{\hat{J}_x}\right)
+K_yK_z\left({\hat{J}_y}{\hat{J}_z}+{\hat{J}_z}{\hat{J}_y}\right).
\end{eqnarray}
Using the above result and squaring the average  $\braket{\hat{n}_4}$ from Eq.~\eqref{eq:n4_avg_fct_Nd_Re-Im_Zd_Jx_Jy_Jz} takes us to the variance given in Eq.~\eqref{eq:n4_Variance_fct_Kd_Zd_J_Schwinger}.

\section{Balanced homodyne detection}
\label{sec:app:BHD}

In order to obtain $\Delta^2\hat{X}_L$, we need to calculate the last two terms from Eq.~\eqref{eq:Delta_2_X_versus_a4}. Using the results from Eq.~\eqref{eq:Kd_Zd_constraints_A_coeffs}, we get
\begin{eqnarray}
\label{eq:a4dagger_a4_minus_average_a4_modsquare}
\Cov\left({\hat{a}_4^\dagger},{\hat{a}_4}\right)
=\frac{1}{2}\left(1+K_z\right)\Cov\left({\hat{a}^\dagger_0},{\hat{a}_0}\right)
+\frac{1}{2}\left(1-K_z\right)\Cov\left({\hat{a}^\dagger_1},{\hat{a}_1}\right)
\nonumber\\
+{ K_x}\Re\left\{\Cov\left({\hat{a}_0},{\hat{a}_1^\dagger}\right)\right\}
-K_y\Im\left\{\Cov\left({\hat{a}_0},{\hat{a}_1^\dagger}\right)\right\}.
\end{eqnarray}
The last term from the variance \eqref{eq:Delta_2_X_versus_a4} is given by
\begin{equation}
\label{eq:Variance_a4}
\Delta^2\hat{a}_4={\mathcal{A}}^2_{40}\Delta^2{\hat{a}_0}
+{\mathcal{A}}^2_{41}\Delta^2{\hat{a}_1}
+2{\mathcal{A}}_{40}{\mathcal{A}}_{41}\Cov({\hat{a}_0},{\hat{a}_1}).
\end{equation}
Please note that the coefficients of the above term are dependent on the phase of local oscillator, $\phi_L$.

\section{Phase sensitivity optimization for a difference-intensity detection scheme}
\label{sec:app:Optimize_diff_det}
The difference intensity detection scheme is described in Section \ref{subsec:diff_intensity_mode} and the corresponding phase sensitivity $\Delta\varphi_{df}$ is given by Eq.~\eqref{eq:Delta_varphi_diff_generic}. In the following we apply the principles discussed in Section \ref{sec:phase_sens_opt_general} in order to obtain both $T'_{opt}$ (or equivalently, $\vartheta'_{opt}$) and $\varphi_{opt}$.

\subsection{Optimum transmission coefficient for the second BS}
\label{subsec:app:Opt_Tprime_diff_mode}

From Eq.~\eqref{eq:del_Delta_varphi_del_vartheta_prime_varphi_equal_ZERO} we get the first constraint
\begin{eqnarray}
\label{eq:vartheta_prime_OPT_diff_det}
\tan{\vartheta'_{opt}}
=\frac{\Delta^2{\hat{J}_y}\sin^2\vartheta
+\Delta^2{\hat{J}_z}\cos^2\vartheta
-\sin2\vartheta\SymCov\left({\hat{J}_y},{\hat{J}_z}\right)}
{\left(\frac{\left(\Delta^2{\hat{J}_z}-\Delta^2{\hat{J}_y}\right)\sin2\vartheta}{2}
+\SymCov\left({\hat{J}_y},{\hat{J}_z}\right)\cos2\vartheta\right){\cos{\varphi_{opt}}}
+\left(\SymCov\left(\hat{J}_x,{\hat{J}_y}\right)\sin\vartheta
-\SymCov\left(\hat{J}_x,{\hat{J}_z}\right)\cos\vartheta\right){\sin{\varphi_{opt}}}},
\quad\quad
\end{eqnarray}
where, as discussed in Appendix \ref{subsec:app:Opt_varphi_diff_mode}, $\varphi_{opt}$ is found in a similar extremization process. Usually though, simplifications can be found by simple arguments leading to a straightforward solution.

Indeed, very often, the working point for a MZI coupled with a difference-intensity detection scheme is given by Eq.~\eqref{eq:app:varphi_OPT_pi_over_2_generic}, we thus have from Eq.~\eqref{eq:vartheta_prime_OPT_diff_det} the simpler formula
\begin{eqnarray}
\label{eq:vartheta_prime_OPT_diff_det_varphi_PI_over_2}
{\vartheta'_{opt}}
=\arctan\left(\frac{\Delta^2{\hat{J}_y}\sin^2\vartheta
+\Delta^2{\hat{J}_z}\cos^2\vartheta
-\sin2\vartheta\SymCov\left({\hat{J}_y},{\hat{J}_z}\right)}
{\SymCov\left({\hat{J}_x},{\hat{J}_y}\right)\sin\vartheta
-\SymCov\left({\hat{J}_x},{\hat{J}_z}\right)\cos\vartheta}
\right).
\end{eqnarray}
Moreover, for many interesting input states  we have $\SymCov\left({\hat{J}_y},{\hat{J}_z}\right)=0=\SymCov\left({\hat{J}_x},{\hat{J}_z}\right)$  (see Tab.~\ref{tab:SmyCov_and_Cov}). Even if it were not the case, the optimization of the first BS via the two parameter QFI, usually results in $\vartheta=\pi/2$ \cite{Ata22}. Thus, Eq.~\eqref{eq:vartheta_prime_OPT_diff_det_varphi_PI_over_2} reduces to 
\begin{eqnarray}
\label{eq:vartheta_prime_OPT_diff_det_BS1_bal_varphi_PI_over_2}
{\vartheta'_{opt}}
=\arctan\left(\frac{\Delta^2{\hat{J}_y}}
{\SymCov\left({\hat{J}_x},{\hat{J}_y}\right)}
\right).
\end{eqnarray}
Now from Tab.~\ref{tab:SmyCov_and_Cov} it is clear that $\SymCov\left({\hat{J}_x},{\hat{J}_y}\right)\neq0$ for a number of input states. However, one must not forget that when looking for the best performance, the optimum input PMC must also be employed. So actually for a coherent plus squeezed vacuum input \eqref{eq:psi_in_coh_plus_sqz_vac}, since $\sin(2\theta_\alpha-\theta)=0$ due to the PMC \eqref{eq:PMC_coh_plus_sqz-vac}, we also get $\SymCov\left(\hat{J}_x,{\hat{J}_y}\right)
=0$. Similarly, for the squeezed-coherent plus squeezed vacuum input state \eqref{eq:psi_in_sqzcoh_plus_sqz_vac} we find 
\begin{equation}
\SymCov\left({\hat{J}_x},{\hat{J}_y}\right)
=-\frac{1}{4}\sinh2r\vert\alpha\vert^2\sin(2\theta_\alpha-\theta)
-\frac{1}{8}{\sinh2r}{\sinh2z}\sin\left({\theta}-{\phi}\right),
\end{equation}
and imposing the PMC \eqref{eq:PMC_sqz-coh_plus_sqz-vac} sets the above symmetrized covariance to zero. For the squeezed-coherent plus squeezed-coherent input state \eqref{eq:psi_in_sqzcoh_plus_sqzcoh} the same conclusions apply for all discussed input PMCs. Applying these arguments to Eq.~\eqref{eq:vartheta_prime_OPT_diff_det_BS1_bal_varphi_PI_over_2} leads to the result
\begin{equation}
\label{eq:app:BS2_opt_balanced_diff_det}
{\vartheta'_{opt}}=\frac{\pi}{2}.
\end{equation}
Even if we assume that $\varphi_{opt}\neq\pi/2$ (but still assuming the first BS balanced \cite{Ata22}), from Eq.~\eqref{eq:vartheta_prime_OPT_diff_det} we have
\begin{eqnarray}
\tan{\vartheta'_{opt}}
=\frac{\Delta^2{\hat{J}_y}}
{-\SymCov\left({\hat{J}_y},{\hat{J}_z}\right){\cos{\varphi_{opt}}}
+\SymCov\left(\hat{J}_x,{\hat{J}_y}\right){\sin{\varphi_{opt}}}}
\end{eqnarray}
and since $\SymCov\left({\hat{J}_y},{\hat{J}_z}\right)=0$ for most of the interesting input states (see Tab.~\ref{tab:SmyCov_and_Cov}), we conclude that the optimum for a difference intensity detection scheme is usually reached if the second beam splitter is balanced (\emph{i. e.} $T'_{opt}=1/\sqrt{2}$).

\subsection{Optimum working point}
\label{subsec:app:Opt_varphi_diff_mode}
The phase sensitivity for a difference intensity detection scheme \emph{i. e.} Eq.~\eqref{eq:Delta_varphi_diff_generic} can be rewritten in the form given by Eq.~\eqref{eq:Delta_varphi_GENERIC_versus_varphi},
where the coefficients are given by
\begin{equation}
\label{eq:app:ABCDEFG_diff_det_coeffs_varphi}
\left\{
\begin{array}{l}
\mathcal{A}_{df} = {\sin^2\vartheta'}\Delta^2{\hat{J}_x}
+\sin^2\vartheta{\cos^2\vartheta'}\Delta^2{\hat{J}_y}
+\cos^2\vartheta{\cos^2\vartheta'}\Delta^2{\hat{J}_z}
-\sin2\vartheta{\cos^2\vartheta'}\SymCov\left({\hat{J}_y},{\hat{J}_z}\right)\\
\mathcal{B}_{df} = -{\sin^2\vartheta'}\Delta^2{\hat{J}_x}
+\cos^2\vartheta{\sin^2\vartheta'}\Delta^2{\hat{J}_y}
+\sin^2\vartheta{\sin^2\vartheta'}\Delta^2{\hat{J}_z}
+\sin2\vartheta{\sin^2\vartheta'}\SymCov\left({\hat{J}_y},{\hat{J}_z}\right)\\
\mathcal{C}_{df} = 
-\left(\cos\vartheta\SymCov\left(\hat{J}_x,{\hat{J}_y}\right)
+\sin\vartheta\SymCov\left(\hat{J}_x,{\hat{J}_z}\right)\right){\sin^2\vartheta'}
 \\
\mathcal{D}_{df} = 
\left(\frac{\sin2\vartheta}{2}\Delta^2{\hat{J}_y}
-\frac{\sin2\vartheta}{2}\Delta^2{\hat{J}_z}
-\cos2\vartheta\SymCov\left({\hat{J}_y},{\hat{J}_z}\right)\right){\sin2\vartheta'}\\
\mathcal{E}_{df} = 
\left(-\sin\vartheta\SymCov\left(\hat{J}_x,{\hat{J}_y}\right)
+\cos\vartheta\SymCov\left(\hat{J}_x,{\hat{J}_z}\right)\right){\sin2\vartheta'}\\
\mathcal{F}_{df} = \braket{\hat{J}_x}{\sin\vartheta'}\\
\mathcal{G}_{df} = \left(\cos\vartheta\braket{\hat{J}_y}+\sin\vartheta\braket{\hat{J}_z}\right){\sin\vartheta'}.
\end{array}
\right.
\end{equation}
If we take no simplifying assumptions, by imposing $\partial\Delta\varphi_{df}/\partial\varphi=0$, as discussed in Appendix \ref{sec:app:varphi_OPT_generic}, we are lead to a 4$^{th}$ degree equation that can be solved either analytically or numerically. Even if we assume the unbalanced case ($\{\vartheta,\vartheta'\}\neq\pi/2$) one of the simple scenarios from Appendix \ref{sec:app:varphi_OPT_generic}  might show up, allowing for a simple solutions for $\varphi_{opt}$.

However, as mentioned previously \cite{Ata22}, the maximization of the two-parameter QFI, usually results in a balanced $BS_1$ (\emph{i. e.} $\vartheta=\pi/2$). In Appendix \ref{subsec:app:Opt_Tprime_diff_mode} we argued that often $BS_2$  ends up being balanced, too, when optimizing the phase sensitivity. Thus, the coefficients from Eq.~\eqref{eq:app:ABCDEFG_diff_det_coeffs_varphi} simplify to
\begin{equation}
\label{eq:app:ABCDEFG_diff_det_coeffs_varphi_balanced}
\left\{
\begin{array}{l}
\mathcal{A}_{df} = \Delta^2{\hat{J}_x}\\
\mathcal{B}_{df} = -\Delta^2{\hat{J}_x}+\Delta^2{\hat{J}_z}\\
\mathcal{C}_{df} = -\SymCov\left(\hat{J}_x,{\hat{J}_z}\right)
 \\
\mathcal{D}_{df} = 0\\
\mathcal{E}_{df} = 0\\
\mathcal{F}_{df} = \braket{\hat{J}_x}\\
\mathcal{G}_{df} = \braket{\hat{J}_z}.
\end{array}
\right.
\end{equation}
For most interesting input states $\SymCov\left(\hat{J}_x,{\hat{J}_z}\right)=0$ (see Tab.~\ref{tab:SmyCov_and_Cov}), thus  $\mathcal{C}_{df} =0$. If $\braket{\hat{J}_x}\neq0$ then the optimum working point is given by \eqref{eq:app:varphi_OPT_frac_A_B_F_G} and for the current discussion (\emph{i. e.} both BS balanced) we have
\begin{equation}
\label{eq:app:varphi_OPT_frac_A_B_F_G_diff}
\varphi_{opt}=\arctan\left(\frac{\braket{\hat{J}_z}\Delta^2{\hat{J}_z}}
{\braket{\hat{J}_x}\Delta^2{\hat{J}_x}}\right)+k\pi,
\end{equation}
with $k\in\mathbb{Z}$. Input states obeying $\braket{\hat{J}_x}=0$ are frequently used (equivalent to at least one of $\braket{\hat{a}_0}$ or $\braket{\hat{a}_1}$ be null), thus $\mathcal{F}_{df}=0$. We conclude that the optimum working point is given by Eq.~\eqref{eq:app:varphi_OPT_pi_over_2_generic} \emph{i. e.}
\begin{equation}
\label{eq:app:varphi_OPT_pi_over_2_diff_det}
\varphi_{opt}=\frac{\pi}{2}+k\pi,
\end{equation}
with $k\in\mathbb{Z}$.

\section{Phase sensitivity optimization for a single mode intensity detection scheme}
\label{sec:app:Optimize_sing_det}
The single mode intensity detection scheme is described in Section \ref{subsec:single_mode}. In the following we apply the extremization process described in Section \ref{sec:phase_sens_opt_general} in order to point out how to compute both  $T'_{opt}$ and $\varphi_{opt}$.

\subsection{Optimum transmission coefficient for the second BS}
\label{subsec:app:Opt_Tprime_single}
In the general case, when optimizing $\vartheta'$ for a non-balanced MZI with a single-mode intensity detection scheme we are led to a 4$^{th}$ degree equation,
\begin{eqnarray}
\label{eq:app:n4_4th_order_equation_vartheta_prime}
(\mathcal{S}_2-\mathcal{S}_0)t^4
+(\mathcal{S}_1 -\mathcal{S}_3)t^3
-4\mathcal{S}_2 t^2
+(\mathcal{S}_1+\mathcal{S}_3)t
+\mathcal{S}_0+\mathcal{S}_2=0,
\end{eqnarray}
where we denoted $\vartheta'=2\arctan t$. The coefficients are given by
\begin{equation}
\label{eq:app:Variance_n4_coeffs_S}
\left\{
\begin{array}{l}
\mathcal{S}_0 = \frac{1}{4}\Delta^2{\hat{N}}
+\Delta^2{\hat{J}_z}\cos^2\vartheta+\Delta^2{\hat{J}_y}\sin^2\vartheta
-\SymCov\left({\hat{J}_y},{\hat{J}_z}\right)\sin2\vartheta\\
\mathcal{S}_1 = 
\left(\left(\Delta^2{\hat{J}_y}-\Delta^2{\hat{J}_z}\right)\sin2\vartheta
-2\SymCov\left({\hat{J}_y},{\hat{J}_z}\right)\cos2\vartheta\right){\cos{\varphi}}\\
\qquad
+2\left(\SymCov\left(\hat{J}_x,{\hat{J}_z}\right)\cos\vartheta
-\SymCov\left(\hat{J}_x,{\hat{J}_y}\right)\sin\vartheta\right){\sin{\varphi}}\\
\mathcal{S}_2 = \Cov\left({\hat{J}_z},{\hat{N}}\right)\cos\vartheta
-\Cov\left({\hat{J}_y},{\hat{N}}\right)\sin\vartheta\\
\mathcal{S}_3 = \Cov\left({\hat{J}_x},{\hat{N}}\right){\sin{\varphi}}
-\left(\Cov\left({\hat{J}_z},{\hat{N}}\right)\sin\vartheta
-\Cov\left({\hat{J}_y},{\hat{N}}\right)\cos\vartheta\right){\cos{\varphi}}\\
\end{array}
\right.
\end{equation}
and at the end of the calculation ${\varphi}$ should be replaced with the value of the working point, ${\varphi_{opt}}$, found in Appendix \ref{subsec:app:Opt_varphi_single}. Since for most input states of interest some of the variances/covariances are null (see Tab.~\ref{tab:SmyCov_and_Cov}), the simpler scenarios described in Appendix \ref{sec:app:varphi_OPT_generic} are possible.

As discussed previously \cite{Ata22}, for most input states, the two-parameter QFI is maximized if $BS_1$ is balanced, \emph{i. e.} $\vartheta=\pi/2$. This remark simplifies the $\mathcal{S}$-coefficients to
\begin{equation}
\label{eq:app:Variance_n4_coeffs_S_BALANCED}
\left\{
\begin{array}{l}
\mathcal{S}_0 = \frac{1}{4}\Delta^2{\hat{N}}+\Delta^2{\hat{J}_y}\\
\mathcal{S}_1 = 
2\SymCov\left({\hat{J}_y},{\hat{J}_z}\right){\cos{\varphi}}
-2\SymCov\left(\hat{J}_x,{\hat{J}_y}\right){\sin{\varphi}}\\
\mathcal{S}_2 = -\Cov\left({\hat{J}_y},{\hat{N}}\right)\\
\mathcal{S}_3 = \Cov\left({\hat{J}_x},{\hat{N}}\right){\sin{\varphi}}
-\Cov\left({\hat{J}_z},{\hat{N}}\right){\cos{\varphi}}.
\end{array}
\right.
\end{equation}
Since for many input states $\SymCov\left({\hat{J}_y},{\hat{J}_z}\right)=0=\Cov\left({\hat{J}_y},{\hat{N}}\right)$ and, moreover, $\SymCov\left(\hat{J}_x,{\hat{J}_y}\right)=0$ if the optimum PMCs are imposed (see Tab.~\ref{tab:SmyCov_and_Cov}), we also have $\mathcal{S}_1=\mathcal{S}_2=0$. Thus, the initial 4$^{th}$ degree Eq.~\eqref{eq:app:n4_4th_order_equation_vartheta_prime} can be rewritten as $\left(\mathcal{S}_0t^2+\mathcal{S}_3t-\mathcal{S}_0\right)\left(t^2-1\right) = 0$,  implying the solution $t=1$ thus
\begin{equation}
\label{eq:app:BS2_opt_balanced_sing_det}
{\vartheta'_{opt}}=\frac{\pi}{2}.
\end{equation}
Remarkably, this solution is independent of the value of the total internal phase shift, $\varphi$.

\subsection{Optimum working point}
\label{subsec:app:Opt_varphi_single}
The phase sensitivity for a single mode intensity detection scheme \emph{i. e.} Eq.~\eqref{eq:Delta_varphi_sing_generic} can also be put in the format given by Eq.~\eqref{eq:Delta_varphi_GENERIC_versus_varphi}, and we have the coefficients 
\begin{equation}
\label{eq:app:ABCDEFG_sing_det_coeffs_varphi}
\left\{
\begin{array}{l}
\mathcal{A}_{sg} = {\sin^2\vartheta'}\Delta^2{\hat{J}_x}
+\sin^2\vartheta{\cos^2\vartheta'}\Delta^2{\hat{J}_y}
+\cos^2\vartheta{\cos^2\vartheta'}\Delta^2{\hat{J}_z}
-\sin2\vartheta{\cos^2\vartheta'}\SymCov\left({\hat{J}_y},{\hat{J}_z}\right)\\
\qquad\qquad
+\frac{1}{4}\Delta^2{\hat{N}}
-\sin\vartheta{\cos\vartheta'}\Cov\left({\hat{J}_y},{\hat{N}}\right)
+\cos\vartheta{\cos\vartheta'}\Cov\left({\hat{J}_z},{\hat{N}}\right)\\
\mathcal{B}_{sg} = -{\sin^2\vartheta'}\Delta^2{\hat{J}_x}
+\cos^2\vartheta{\sin^2\vartheta'}\Delta^2{\hat{J}_y}
+\sin^2\vartheta\sin^2\vartheta'\Delta^2{\hat{J}_z}
+\sin2\vartheta{\sin^2\vartheta'}\SymCov\left({\hat{J}_y},{\hat{J}_z}\right)\\
\mathcal{C}_{sg} = 
-\cos\vartheta{\sin^2\vartheta'}\SymCov\left(\hat{J}_x,{\hat{J}_y}\right)
-\sin\vartheta{\sin^2\vartheta'}\SymCov\left(\hat{J}_x,{\hat{J}_z}\right)
 \\
\mathcal{D}_{sg} = 
\frac{1}{2}\sin2\vartheta{\sin2\vartheta'}\Delta^2{\hat{J}_y}
-\frac{1}{2}\sin2\vartheta{\sin2\vartheta'}\Delta^2{\hat{J}_z}
-\cos2\vartheta{\sin2\vartheta'}\SymCov\left({\hat{J}_y},{\hat{J}_z}\right)\\
\qquad\qquad
-\cos\vartheta{\sin\vartheta'}\Cov\left({\hat{J}_y},{\hat{N}}\right)
-\sin\vartheta{\sin\vartheta'}\Cov\left({\hat{J}_z},{\hat{N}}\right)\\
\mathcal{E}_{sg} = 
-\sin\vartheta{\sin2\vartheta'}\SymCov\left(\hat{J}_x,{\hat{J}_y}\right)
+\cos\vartheta{\sin2\vartheta'}\SymCov\left(\hat{J}_x,{\hat{J}_z}\right)
+{\sin\vartheta'}\Cov\left({\hat{J}_x},{\hat{N}}\right)\\
\mathcal{F}_{sg} = \braket{\hat{J}_x}{\sin\vartheta'}\\
\mathcal{G}_{sg} = \left(\cos\vartheta\braket{\hat{J}_y}+\sin\vartheta\braket{\hat{J}_z}\right){\sin\vartheta'}.
\end{array}
\right.
\end{equation}
Following the discussion from Appendix \ref{sec:app:varphi_OPT_generic}, the extremization process in the generic case with an unbalanced MZI results an a 4$^{th}$ degree equation. However, as already discussed, the phase sensitivity optimization for a single-mode intensity detection scheme results in a balanced MZI ($BS_1$ is usually balanced from the maximization of the two-parameter QFI and $BS_2$ is discussed in Appendix \ref{subsec:app:Opt_Tprime_single}). This implies the much simpler coefficients:
\begin{equation}
\label{eq:app:ABCDEFG_sing_det_coeffs_varphi_BALANCED}
\left\{
\begin{array}{l}
\mathcal{A}_{sg} = \Delta^2{\hat{J}_x}
+\frac{1}{4}\Delta^2{\hat{N}}\\
\mathcal{B}_{sg} = -\Delta^2{\hat{J}_x}+\Delta^2{\hat{J}_z}\\
\mathcal{C}_{sg} = -\SymCov\left(\hat{J}_x,{\hat{J}_z}\right)
 \\
\mathcal{D}_{sg} = -\Cov\left({\hat{J}_z},{\hat{N}}\right)\\
\mathcal{E}_{sg} = \Cov\left({\hat{J}_x},{\hat{N}}\right)\\
\mathcal{F}_{sg} = \braket{\hat{J}_x}\\
\mathcal{G}_{sg} = \braket{\hat{J}_z}.
\end{array}
\right.
\end{equation}
As discussed before, for almost all considered input states $\SymCov\left(\hat{J}_x,{\hat{J}_z}\right)=0=\Cov\left({\hat{J}_x},{\hat{N}}\right)$, thus $\mathcal{C}_{sg}=\mathcal{E}_{sg}=0$. Moreover, if $\mathcal{F}_{sg}=0$, we are able to apply Eq.~\eqref{eq:varphi_OPT_frac_A_B_D} and obtain the working point
\begin{equation}
\label{eq:varphi_OPT_frac_A_B_D_single}
{ \varphi_{opt}}=\pm2\arctan\sqrt[4]{\frac{
\Delta^2{\hat{J}_z}+\frac{1}{4}\Delta^2{\hat{N}}-\Cov\left({\hat{J}_z},{\hat{N}}\right)}
{\Delta^2{\hat{J}_z}+\frac{1}{4}\Delta^2{\hat{N}}+\Cov\left({\hat{J}_z},{\hat{N}}\right)}}+2k\pi.
\end{equation}
For example, applying the above result to the coherent plus squeezed vacuum input \eqref{eq:psi_in_coh_plus_sqz_vac} yields \cite{API18}
\begin{equation}
\label{eq:varphi_opt_coh_plus_sqz_vac_sing_det}
\varphi_{opt}=\pm2\arctan\sqrt{\frac{\sqrt{2}\vert\alpha\vert}{\sinh2r}}+2k\pi.
\end{equation}
Applying Eq.~\eqref{eq:varphi_OPT_frac_A_B_D_single} to a squeezed-coherent plus squeezed vacuum input Eq.~\eqref{eq:psi_in_sqzcoh_plus_sqz_vac} yields \cite{Ata19}
\begin{equation}
\label{eq:varphi_opt_sqz-coh_plus_sqz_vac_sing_det}
\varphi_\textrm{opt}
=\pm2\arctan\sqrt[4]{\frac{{\sinh^22z}
+2\vert\alpha\vert^2\left({\cosh2z}-{\sinh2z}\cos(2\theta_\alpha-\phi)\right)}
{{\sinh^22r}}}.
\end{equation}

\section{Phase sensitivity optimization for a balanced homodyne detection scheme}
\label{sec:app:Optimum_BHD}

The balanced homodyne detection scheme is described in Section \ref{subsec:BHD_scheme}. In the following we apply the extremization process described in Section \ref{sec:phase_sens_opt_general} in order to point out how to compute both  $\vartheta'_{opt}$ and $\varphi_{opt}$. The specificity of this state is the existence of the phase of the local oscillator. As described below, this phase is typically matched with the one of the input laser \eqref{eq:app:phi_L_equal_theta_alpha}.

\subsection{Optimum transmission coefficient for the second BS}
\label{subsec:app:Opt_Tprime_homodyne}

Assuming the convention from Eq.~\eqref{eq:one_phase_shift_CONVENTION}, the phase sensitivity for a BHD scheme is described in Section \ref{subsec:diff_intensity_mode} and given by Eq.~\eqref{eq:Delta_varphi_hom_1phase_generic}. Applying the principles from Eq.~\eqref{eq:del_Delta_varphi_del_vartheta_prime_varphi_equal_ZERO} to this case yields the optimum $BS_2$ transmission coefficient
\begin{eqnarray}
{\vartheta'}=2\arctan\frac{L_{bhd}}{U_{bhd}},
\end{eqnarray}
where we have the terms
\begin{eqnarray}
\label{eq:Upper_BHD_one_phase}
U_{bhd}=
-\sin\vartheta{\cos{\varphi}}\varpi^-
-\sin{\vartheta}\Re\left\{e^{-i(2\phi_L+\varphi)}\left(\Delta^2{\hat{a}_0}+\Delta^2{\hat{a}_1}\right)\right\}
+2{\sin{\varphi}}\Re\left\{\Cov\left({\hat{a}_0},{\hat{a}_1^\dagger}\right)\right\}
\nonumber\\
-2\cos\vartheta{\cos{\varphi}}\Re\left\{i\Cov\left({\hat{a}_0},{\hat{a}_1^\dagger}\right)\right\}
+2\cos{\vartheta}\Re\left\{e^{-i(2\phi_L+\varphi)}i
\Cov({\hat{a}_0},{\hat{a}_1})\right\}
\end{eqnarray}
and
\begin{eqnarray}
\label{eq:Lower_BHD_one_phase}
L_{bhd} = -(1+{\varpi^+})
-\Re\left\{e^{-i2\phi_L}\left(\Delta^2{\hat{a}_0}-\Delta^2{\hat{a}_1}\right)\right\}
-\cos\vartheta\varpi^-
-\cos{\vartheta}\Re\left\{e^{-i2\phi_L}\left(\Delta^2{\hat{a}_0}
+\Delta^2{\hat{a}_1}\right)\right\}
\nonumber\\
+2\sin\vartheta\Re\left\{i\Cov\left({\hat{a}_0},{\hat{a}_1^\dagger}\right)\right\}
-2\Re\left\{ie^{-i2\phi_L}\sin\vartheta\Cov({\hat{a}_0},{\hat{a}_1})\right\}.
\end{eqnarray}
For a more compact writing, we introduce the following notations:
\begin{equation}
\label{eq:Varpi_Varpi_PLUS_notations}
\left\{
\begin{array}{l}
\varpi^-
=\Cov\left({\hat{a}_0^\dagger},{\hat{a}_0}\right)
-\Cov\left({\hat{a}_1^\dagger},{\hat{a}_1}\right)\\
\varpi^+ 
=\Cov\left({\hat{a}_0^\dagger},{\hat{a}_0}\right)
+\Cov\left({\hat{a}_1^\dagger},{\hat{a}_1}\right).
\end{array}
\right.
\end{equation}
If we assume a non-entangled input state $\Cov({\hat{a}_0},{\hat{a}_1})=0$, we thus get the simpler solution
\begin{eqnarray}
\label{eq:tan_vartheta_prime_OPT_non-entangled}
{\vartheta'}_{opt}
=2\arctan\left(\frac{1+{\varpi^+}
+\cos\vartheta\varpi
+\Re\left\{e^{-i2\phi_L}\left(\Delta^2{\hat{a}_0}-\Delta^2{\hat{a}_1})
\right)\right\}
+\cos\vartheta\Re\left\{e^{-i2\phi_L}\left(\Delta^2{\hat{a}_0}+\Delta^2{\hat{a}_1})
\right)\right\}
}
{\sin\vartheta\left({\cos{\varphi}}\varpi^-
+\Re\left\{e^{-i(2\phi_L+\varphi)}\left(\Delta^2{\hat{a}_0}
+\Delta^2{\hat{a}_1}\right)\right\}\right)}
\right).
\end{eqnarray}
However, quite often, the optimum working point for a balanced homodyne detection scheme is given by Eq.~\eqref{eq:app:varphi_OPT_pi_over_2_homodyne}. We thus have the simpler solution
\begin{eqnarray}
\label{eq:tan_vartheta_prime_OPT_non-entangled_at_varphi_OPT}
\vartheta'_{opt}
=2\arctan\left(\frac{1+{\varpi^+}+\cos\vartheta\varpi
+\Re\left\{e^{-i2\phi_L}\left(\Delta^2{\hat{a}_0}-\Delta^2{\hat{a}_1})
\right)\right\}
+\cos\vartheta\Re\left\{e^{-i2\phi_L}\left(\Delta^2{\hat{a}_0}+\Delta^2{\hat{a}_1})
\right)\right\}
}
{\sin\vartheta\left(-\varpi^-
-\Re\left\{e^{-i2\phi_L}\left(\Delta^2{\hat{a}_0}
+\Delta^2{\hat{a}_1}\right)\right\}\right)}
\right).
\end{eqnarray}
For example, by applying the above equation to the coherent plus squeezed vacuum input state \eqref{eq:psi_in_coh_plus_sqz_vac}, if we set the phase of the local oscillator to \eqref{eq:app:phi_L_equal_theta_alpha} and impose the optimum PMC \eqref{eq:PMC_coh_plus_sqz-vac} we get
\begin{equation}
\label{eq:Tprime_opt_coh_sqz_vac_vartheta}
{T'}^{(i)}_{opt}=\cos\frac{\vartheta'_{opt}}{2}
=\frac{\cos\frac{\vartheta}{2}\sin\frac{\vartheta}{2}(1-e^{-2r})}
{\sqrt{1-\cos^2\frac{\vartheta}{2}\left(1-e^{-4r}\right)}}
\end{equation}
and by replacing $\cos\frac{\vartheta}{2}$ with ${T}^{(i)}_{opt}$ we recover equation (81) from reference \cite{Ata20}.

\subsection{Optimum working point}
\label{subsec:app:Opt_varphi_homodyne}
Similar to the previously discussed detection schemes, we can write the phase sensitivity in the form of Eq.~\eqref{eq:Delta_varphi_GENERIC_versus_varphi}, where the coefficients are given by
\begin{equation}
\label{eq:app:ABCDEFG_homodyne_coeffs_varphi}
\left\{
\begin{array}{l}
\mathcal{A}_{hom} = \frac{1}{4}
+\frac{1}{4}\left(\Cov({\hat{a}_0^\dagger},{\hat{a}_0})+\Cov({\hat{a}_1^\dagger},{\hat{a}_1})\right)
+\frac{\cos\vartheta{\cos\vartheta'}}{4}\left(\Cov({\hat{a}_0^\dagger},{\hat{a}_0})-\Cov({\hat{a}_1^\dagger},{\hat{a}_1})\right)
\\
\qquad\quad+\frac{1}{2}\left(\cos^2\frac{\vartheta}{2}\cos^2\frac{\vartheta'}{2}
-\sin^2\frac{\vartheta}{2}\sin^2\frac{\vartheta'}{2}\right)
\Re\left\{\Delta^2{\hat{a}_0}\right\}
-\frac{1}{4}\left(1-\cos{\vartheta}\cos{\vartheta'}\right)\Re\left\{\Delta^2{\hat{a}_1}\right\}
\\
\qquad\quad+\frac{1}{2}\sin\vartheta{\cos\vartheta'}\left(\Im\left\{\Cov\left({\hat{a}_0},{\hat{a}_1^\dagger}\right)\right\}
-\Im\left\{\Cov({\hat{a}_0},{\hat{a}_1})\right\}
\right)\\
\mathcal{B}_{hom} = \left(\sin^2\frac{\vartheta}{2}\Re\left\{\Delta^2{\hat{a}_0}\right\}
-\cos^2\frac{\vartheta}{2}\Re\left\{\Delta^2{\hat{a}_1}\right\}
+\sin{\vartheta}\Im\left\{\Cov({\hat{a}_0},{\hat{a}_1})\right\}
\right)\sin^2\frac{\vartheta'}{2}\\
\mathcal{C}_{hom} = \frac{1}{2}\left(\sin^2\frac{\vartheta}{2}\Im\left\{\Delta^2{\hat{a}_0}\right\}
-\cos^2\frac{\vartheta}{2}\Im\left\{\Delta^2{\hat{a}_1}\right\}
-\sin{\vartheta}\Re\left\{\Cov({\hat{a}_0},{\hat{a}_1})\right\}
\right)\sin^2\frac{\vartheta'}{2}\\
\mathcal{D}_{hom} = \frac{1}{2}\left(\cos\vartheta\left(\Im\left\{\Cov\left({\hat{a}_0},{\hat{a}_1^\dagger}\right)\right\}
-\Im\left\{\Cov({\hat{a}_0},{\hat{a}_1})\right\}\right)\right.\\
\quad\qquad\left.-\frac{\sin\vartheta}{2}\left(\Cov({\hat{a}_0^\dagger},{\hat{a}_0})-\Cov({\hat{a}_1^\dagger},{\hat{a}_1})+\Re\left\{\Delta^2{\hat{a}_0}\right\}
+\Re\left\{\Delta^2{\hat{a}_1}\right\}\right)\right){\sin\vartheta'}\\
\mathcal{E}_{hom} = \frac{1}{2}\left(\Re\left\{\Cov\left({\hat{a}_0},{\hat{a}_1^\dagger}\right)\right\}
-\frac{1}{2}\sin{\vartheta}\left(\Im\left\{\Delta^2{\hat{a}_0}\right\}+\Im\left\{\Delta^2{\hat{a}_1}\right\}\right)
+\cos{\vartheta}\Re\left\{\Cov({\hat{a}_0},{\hat{a}_1})\right\}
\right){\sin\vartheta'}\\
\mathcal{F}_{hom} = \left(-\sin\frac{\vartheta}{2}\Im\left\{\braket{\hat{a}_0}\right\}
+\cos\frac{\vartheta}{2}\Re\left\{\braket{\hat{a}_1}\right\}\right)\sin\frac{\vartheta'}{2}\\
\mathcal{G}_{hom} = \left(\sin\frac{\vartheta}{2}\Re\left\{\braket{\hat{a}_0}\right\}
+\cos\frac{\vartheta}{2}\Im\left\{\braket{\hat{a}_1}\right\}\right)\sin\frac{\vartheta'}{2}
\end{array}
\right.
\end{equation}
and we set $\phi_L=0$ for readability. If we assume a non-entangled input state (and restore $\phi_L$), the coefficients are expressed by
\begin{equation}
\label{eq:app:ABCDEFG_homodyne_coeffs_varphi_NON_entangled}
\left\{
\begin{array}{l}
\mathcal{A}_{hom} = \frac{1}{4}
+\frac{1}{4}\left(\Cov({\hat{a}_0^\dagger},{\hat{a}_0})+\Cov({\hat{a}_1^\dagger},{\hat{a}_1})\right)
+\frac{\cos\vartheta{\cos\vartheta'}}{4}\left(\Cov({\hat{a}_0^\dagger},{\hat{a}_0})-\Cov({\hat{a}_1^\dagger},{\hat{a}_1})\right)\\
+\frac{1}{2}\left(\cos^2\frac{\vartheta}{2}\cos^2\frac{\vartheta'}{2}
-\sin^2\frac{\vartheta}{2}\sin^2\frac{\vartheta'}{2}\right)
\Re\left\{e^{-i2\phi_L}\Delta^2{\hat{a}_0}\right\}
-\frac{1}{2}\left(\sin^2\frac{\vartheta}{2}\cos^2\frac{\vartheta'}{2}
+\cos^2\frac{\vartheta}{2}\sin^2\frac{\vartheta'}{2}\right)\Re\left\{e^{-i2\phi_L}\Delta^2{\hat{a}_1}\right\}\\
\mathcal{B}_{hom} = \left(\sin^2\frac{\vartheta}{2}\Re\left\{e^{-i2\phi_L}\Delta^2{\hat{a}_0}\right\}
-\cos^2\frac{\vartheta}{2}\Re\left\{e^{-i2\phi_L}\Delta^2{\hat{a}_1}\right\}
\right)\sin^2\frac{\vartheta'}{2}\\
\mathcal{C}_{hom} = \frac{1}{2}\left(\sin^2\frac{\vartheta}{2}\Im\left\{e^{-2\phi_L}\Delta^2{\hat{a}_0}\right\}
-\cos^2\frac{\vartheta}{2}\Im\left\{e^{-2\phi_L}\Delta^2{\hat{a}_1}\right\}
\right)\sin^2\frac{\vartheta'}{2}\\
\mathcal{D}_{hom} = -\frac{1}{4}\left(\Cov({\hat{a}_0^\dagger},{\hat{a}_0})
-\Cov({\hat{a}_1^\dagger},{\hat{a}_1})+\Re\left\{e^{-i2\phi_L}\Delta^2{\hat{a}_0}\right\}
+\Re\left\{e^{-i2\phi_L}\Delta^2{\hat{a}_1}\right\}\right)\sin\vartheta{\sin\vartheta'}\\
\mathcal{E}_{hom} = -\frac{1}{4}\left(
\Im\left\{e^{-i2\phi_L}\Delta^2{\hat{a}_0}\right\}
+\Im\left\{e^{-i2\phi_L}\Delta^2{\hat{a}_1}\right\}
\right)\sin{\vartheta}{\sin\vartheta'}\\
\mathcal{F}_{hom} = \left(-\sin\frac{\vartheta}{2}\Im\left\{e^{-\phi_L}\braket{\hat{a}_0}\right\}
+\cos\frac{\vartheta}{2}\Re\left\{e^{-\phi_L}\braket{\hat{a}_1}\right\}\right)\sin\frac{\vartheta'}{2}\\
\mathcal{G}_{hom} = \left(\sin\frac{\vartheta}{2}\Re\left\{e^{-\phi_L}\braket{\hat{a}_0}\right\}
+\cos\frac{\vartheta}{2}\Im\left\{e^{-\phi_L}\braket{\hat{a}_1}\right\}\right)\sin\frac{\vartheta'}{2}.
\end{array}
\right.
\end{equation}
In the general case, the optimum working point is found by solving Eq.~\eqref{eq:app:varphi_OPT_4th_degree_generic}. However, quite often, when using the optimum input PMC, it turns out that $\mathcal{C}_{hom}=\mathcal{E}_{hom}=\mathcal{G}_{hom}=0$ and we can use the result \eqref{eq:app:varphi_OPT_pi_generic} for the optimum working point. For most input states, further refinements allow one to show that
\begin{equation}
\label{eq:app:varphi_OPT_pi_over_2_homodyne}
\varphi_{opt}=\pi+2k\pi
\end{equation}
with $k\in\mathbb{Z}$.

We would like to show how the optimum input PMC is connected to the constraint $\mathcal{C}_{hom}=\mathcal{E}_{hom}=\mathcal{G}_{hom}=0$. Let us assume that we have the coherent plus squeezed vacuum input state given in Eq.~\eqref{eq:psi_in_sqzcoh_plus_sqz_vac}. (The argument is similar for other input states.) It is well known that for optimal performance one must match the local oscillator's phase with the input coherent phase, \emph{i. e.} 
\begin{equation}
\label{eq:app:phi_L_equal_theta_alpha}
\phi_L=\theta_\alpha. 
\end{equation}
Thus, the nonzero term of $\mathcal{G}_{hom}$ ($\braket{\hat{a}_0}=0$ for the input state considered) reads
\begin{equation}
\Im\left\{e^{-\phi_L}\braket{\hat{a}_1}\right\}=
\vert\alpha\vert\Im\left\{e^{-(\phi_L-\theta_\alpha)}\right\}
\end{equation}
and $\Im\left\{e^{-\phi_L}\braket{\hat{a}_1}\right\}=0$ because of Eq.~\eqref{eq:app:phi_L_equal_theta_alpha}. When it comes to $\mathcal{C}_{hom}$ and $\mathcal{E}_{hom}$, we seem to have the non-zero terms
\begin{equation}
\Im\left\{e^{-2\phi_L}\Delta^2{\hat{a}_0}\right\}
=\Im\left\{e^{-2\theta_\alpha}\Delta^2{\hat{a}_0}\right\}
=-\frac{1}{2}\sinh2r\Im\left\{e^{-(2\theta_\alpha-\theta)}\right\}
\end{equation}
and
\begin{equation}
\Im\left\{e^{-2\phi_L}\Delta^2{\hat{a}_1}\right\}
=\Im\left\{e^{-2\theta_\alpha}\Delta^2{\hat{a}_1}\right\}
=-\frac{1}{2}\sinh2z\Im\left\{e^{-(2\theta_\alpha-\phi)}\right\}.
\end{equation}
However, by imposing the optimum input PMC \eqref{eq:PMC_sqz-coh_plus_sqz-vac} guarantees that these terms vanish, too.

\section{Generic optimum working point calculation}
\label{sec:app:varphi_OPT_generic}
All detection schemes discussed in Section \ref{sec:detection_schemes} yield the same structure of the  phase sensitivity when it comes to the $\varphi$-dependence, namely
\begin{equation}
\label{eq:Delta_varphi_GENERIC_versus_varphi}
\Delta\varphi=\frac{\sqrt{\mathcal{A}+\mathcal{B}{\cos^2{\varphi}}
+\mathcal{C}{\sin{2\varphi}}+\mathcal{D}{\cos{\varphi}}
+\mathcal{E}{\sin{\varphi}}}}
{\vert\mathcal{F}\cos{\varphi}+\mathcal{G}\sin{\varphi}\vert},
\end{equation}
where the $\mathcal{A}\ldots\mathcal{G}$ coefficients are specific to each respective detection scheme and are given in Appendices \ref{subsec:app:Opt_varphi_diff_mode}, \ref{subsec:app:Opt_varphi_single} and respectively \ref{subsec:app:Opt_varphi_homodyne}. In order to find the working point $\varphi_{opt}$ for a generic detection scheme, we start from Eq.~\eqref{eq:Delta_varphi_GENERIC_versus_varphi} and impose $\partial_\varphi\Delta\varphi=0$. For the most general case (assuming that none of the coefficients is zero), we are led to the trigonometric equation
\begin{eqnarray}
\left(\mathcal{D}\mathcal{F}-\mathcal{E}\mathcal{G}\right)\sin{\varphi}{\cos{\varphi}}
+(2\mathcal{E}\mathcal{F}-\mathcal{D}\mathcal{G}){\sin^2{\varphi}}
+(\mathcal{E}\mathcal{F}-2\mathcal{D}\mathcal{G}){\cos^2{\varphi}}
\nonumber\\
+2\left(\mathcal{C}\mathcal{F}-\mathcal{B}\mathcal{G}-\mathcal{A}\mathcal{G}\right)\cos{\varphi}
+2\left(\mathcal{A}\mathcal{F}-\mathcal{C}\mathcal{G}\right){\sin{\varphi}}
=0.
\end{eqnarray}
Using the parametrization $\cos\varphi=\frac{1-t^2}{1+t^2}$ and $\sin\varphi=\frac{2t}{1+t^2}$ we are led to a 4$^{th}$ degree equation,
\begin{eqnarray}
\label{eq:app:varphi_OPT_4th_degree_generic}
\left[\mathcal{E}\mathcal{F}-2\mathcal{D}\mathcal{G}
-2\mathcal{C}\mathcal{F}+2(\mathcal{A}+\mathcal{B})\mathcal{G}\right]{t^4}
+\left[4\left(\mathcal{A}\mathcal{F}-\mathcal{C}\mathcal{G}\right)
-2\left(\mathcal{D}\mathcal{F}-\mathcal{E}\mathcal{G}\right)\right]{ t^3}
+6\mathcal{E}\mathcal{F}{ t^2}
\nonumber\\
+\left[2\left(\mathcal{D}\mathcal{F}-\mathcal{E}\mathcal{G}\right)
+4\left(\mathcal{A}\mathcal{F}-\mathcal{C}\mathcal{G}\right)\right]{ t}
+2\left[\mathcal{C}\mathcal{F}-(\mathcal{A}+\mathcal{B})\mathcal{G}\right]
+(\mathcal{E}\mathcal{F}-2\mathcal{D}\mathcal{G})
=0
\end{eqnarray}
that can be solved either analytically or numerically. The optimum internal phase shift is found among the solutions $t_{sol}\in\mathbb{R}$ of Eq.~\eqref{eq:app:varphi_OPT_4th_degree_generic} and then we have $\varphi_{opt}=2\arctan t_{sol}$.

However, quite often simplifications are found. We discuss below the scenarios that present an interest for the current analysis.

\begin{enumerate}
	\item[i)] If $\mathcal{C}=\mathcal{E}=\mathcal{F}=0$ then the optimum working point is given by
\begin{equation}
\label{eq:varphi_OPT_frac_A_B_D}
{ \varphi_{opt}}=\pm2\arctan\sqrt[4]{\frac{\mathcal{A}+\mathcal{B}+\mathcal{D}}{\mathcal{A}+\mathcal{B}-\mathcal{D}}}+2k\pi
\end{equation}
with $k\in\mathbb{Z}$.
	\item[ii)] If $\mathcal{C}=\mathcal{E}=\mathcal{F}=0$ and also $\mathcal{D}=0$ then the optimum is given by
\begin{equation}
\label{eq:app:varphi_OPT_pi_over_2_generic}
\varphi_{opt}=\frac{\pi}{2}+k\pi
\end{equation}
with $k\in\mathbb{Z}$.
	\item[iii)] If $\mathcal{C}=\mathcal{E}=\mathcal{G}=0$ then the optimum is found among the solutions
\begin{equation}
\label{eq:app:varphi_OPT_pi_generic}
\varphi_{opt}=k\pi
\end{equation}
with $k\in\mathbb{Z}$.
	\item[iv)] If $\mathcal{C}=\mathcal{D}=\mathcal{E}=0$ then the optimum is found among the solutions
\begin{equation}
\label{eq:app:varphi_OPT_frac_A_B_F_G}
\varphi_{opt}=\arctan\left(\frac{(\mathcal{A}+\mathcal{B})\mathcal{G}}{\mathcal{A}\mathcal{F}}\right)+k\pi
\end{equation}
with $k\in\mathbb{Z}$.
\end{enumerate}

\begin{table}
\begin{tabular}{|c|c|c|c|c|c|}
\hline
input & single & double   
& coherent plus
& squeezed-coherent  
& coherent\\
state & coherent & coherent  
& squeezed vacuum
& plus squeezed vacuum 
& plus Fock\\
$\ket{\psi_{in}}$ 
& $\ket{\alpha_10_0}$ 
& $\ket{\alpha_1\beta_0}$  
& $\ket{\alpha_1\xi_0}$
& $\ket{(\alpha\zeta)_1\xi_0}$
& $\ket{\alpha_1n_0}$\\
\hline
$\SymCov\left({\hat{J}_x},{\hat{J}_y}\right)$ 
& 0 
& 0
& $-\frac{\sinh2r\vert\alpha\vert^2\sin(2\theta_\alpha-\theta)}{4}$ 
& $-\frac{\sinh2r\vert\alpha\vert^2\sin(2\theta_\alpha-\theta)}{4}-\frac{{\sinh2r}{\sinh2z}\sin\left({\theta}
-{\phi}\right)}{8}$  
& 0\\
\hline
$\SymCov\left({\hat{J}_x},{\hat{J}_z}\right)$  
& 0 
& 0  
& 0 
& 0 
& 0\\
\hline
$\SymCov\left({\hat{J}_y},{\hat{J}_z}\right)$  
& 0 
& 0 
& 0 
& 0 
& 0\\
\hline
$\Cov\left({\hat{J}_x},{\hat{N}}\right)$  
& 0 
& $\vert\alpha\beta\vert\cos(\theta_\alpha-\theta_\beta)$ 
& 0 
& 0 
& 0\\
\hline
$\Cov\left({\hat{J}_y},{\hat{N}}\right)$  
& 0 
& $\vert\alpha\beta\vert\sin(\theta_\alpha-\theta_\beta)$ 
& 0  
& 0 
& 0\\
\hline
$\Cov\left({\hat{J}_z},{\hat{N}}\right)$  
& $-\frac{1}{2}\vert\alpha\vert^2$ 
& $\frac{1}{2}\left(\vert\beta\vert^2-\vert\alpha\vert^2\right)$  
& $\frac{\sinh^22r}{2}-\vert\alpha\vert^2$
& $\frac{\sinh^22r-\sinh^22z}{2}-\vert\alpha\vert^2(\cosh2z-\sinh2z\cos(2\theta_\alpha-\phi))$
& $-\frac{1}{2}\vert\alpha\vert^2$\\
\hline
\end{tabular}
\caption{The covariances and symmetrized covariances for a number of input states.}
\label{tab:SmyCov_and_Cov}
\end{table}

\section{Calculation for some needed variances and covariances}
Throughout the main part of this paper we referred to some variances, covariances and symmetrized covariances for the various input states considered. They are given in Tab.~\ref{tab:SmyCov_and_Cov} and \ref{tab:SmyCov_and_Cov_sqzcoh_sqzcoh}. In the following, we detail some computational details needed to obtain the aforementioned results.

For the squeezed-coherent plus squeezed-coherent input state \eqref{eq:psi_in_sqzcoh_plus_sqzcoh}, we find the variances of the Schwinger pseudo-angular momentum operators
\begin{eqnarray}
\label{eq:app:Variance_Jx_sqzcoh_sqzcoh}
\Delta^2{\hat{J}_x}
=\frac{1}{4}\Big(
{\vert\beta\vert^2}\left(\cosh2z-\sinh2z\cos(2\theta_\beta-\phi)\right)
+{\vert\alpha\vert^2}\left({\cosh2r}-\sinh2r\cos(2\theta_\alpha-\theta)\right)
\nonumber\\
+\frac{1}{2}\left({\cosh2r}\cosh2z
+\sinh2r\sinh2z\cos(\theta-\phi)-1\right)
\Big),
\end{eqnarray}
\begin{eqnarray}
\label{eq:app:Variance_Jy_sqzcoh_sqzcoh}
\Delta^2{\hat{J}_y}
=\frac{1}{4}\Big(
{\vert\beta\vert^2}\left(\cosh2z+\sinh2z\cos(2\theta_\beta-\phi)\right)
+{\vert\alpha\vert^2}\left({\cosh2r}+\sinh2r\cos(2\theta_\alpha-\theta)\right)
\nonumber\\
+\frac{1}{2}\left({\cosh2r}\cosh2z-\sinh2z\sinh2r\cos(\theta-\phi)-1\right)
\Big),
\end{eqnarray}
and
\begin{equation}
\label{eq:app:Variance_Jz_sqzcoh_sqzcoh}
\Delta^2{\hat{J}_z}
=\frac{1}{4}\left(\frac{\sinh^22r}{2}
+\vert\beta\vert^2\left(\cosh2r-\sinh2r\cos(2\theta_\beta-\theta)\right)
-\frac{\sinh^22z}{2}-\vert\alpha\vert^2\left(\cosh2z-\sinh2z\cos\left(2\theta_\alpha-\phi\right)\right)
\right).
\end{equation}
The symmetrized covariances are found to be
\begin{equation}
\label{eq:app:SymCov_Jx_Jy_sqzcoh_sqzcoh}
\SymCov\left({\hat{J}_x},{\hat{J}_y}\right)
=-\frac{1}{8}{\sinh2r}{\sinh2z}\sin(\theta-\phi)
-\frac{1}{4}\vert{\alpha}\vert^2{\sinh2r}\sin(2\theta_\alpha-\theta)
+\frac{1}{4}\vert{\beta}\vert^2{\sinh2z}\sin(2\theta_\beta-\phi),
\end{equation}
\begin{equation}
\label{eq:app:SymCov_Jx_Jz_sqzcoh_sqzcoh}
\SymCov\left({\hat{J}_x},{\hat{J}_z}\right)
=\frac{1}{2}\vert\alpha\beta\vert(\sinh^2r-\sinh^2z)\cos(\theta_\alpha-\theta_\beta)
-\frac{\vert\alpha\beta\vert}{4}\left(\sinh2r\cos(\theta_\alpha+\theta_\beta-\theta)
-\sinh2z\cos(\theta_\alpha+\theta_\beta-\phi)\right),
\end{equation}
and
\begin{equation}
\label{eq:app:SymCov_Jy_Jz_sqzcoh_sqzcoh}
\SymCov\left({\hat{J}_y},{\hat{J}_z}\right)
=\frac{1}{2}\vert\alpha\beta\vert(\sinh^2r-\sinh^2z)\sin(\theta_\alpha-\theta_\beta)
-\frac{\vert\alpha\beta\vert}{4}\left(\sinh2r\sin(\theta_\alpha+\theta_\beta-\theta)
+\sinh2z\sin(\theta_\alpha+\theta_\beta-\phi)\right).
\end{equation}
The variance of the total input photon number operator is found to be
\begin{eqnarray}
\label{eq:app:Variance_N_sqzcoh_sqzcoh}
\Delta^2{\hat{N}}
=\frac{\sinh^22r}{2}+\frac{\sinh^22z}{2}
+\vert\beta\vert^2\left(\cosh2r-\sinh2r\cos(2\theta_\beta-\theta)\right)
+\vert\alpha\vert^2\left(\cosh2z-\sinh2z\cos\left(2\theta_\alpha-\phi\right)\right)
\end{eqnarray}
and finally the covariances of the $\hat{J}$-operators with the input photon number operator $\hat{N}$ are
\begin{equation}
\Cov\left({\hat{J}_x},{\hat{N}}\right)
=\vert\alpha\beta\vert(\sinh^2r+\sinh^2z+1)\cos(\theta_\alpha-\theta_\beta)
-\frac{\vert\alpha\beta\vert}{2}\left(\sinh2r\cos(\theta_\alpha+\theta_\beta-\theta)
+\sinh2z\cos(\theta_\alpha+\theta_\beta-\phi)\right),
\end{equation}
\begin{equation}
\Cov\left({\hat{J}_y},{\hat{N}}\right)
=\vert\alpha\beta\vert\left(\sinh^2r+\sinh^2z+1\right)\sin(\theta_\alpha-\theta_\beta)
-\frac{\vert\alpha\beta\vert}{2}\left(\sinh2r\sin(\theta_\alpha+\theta_\beta-\theta)
-\sinh2z\sin(\theta_\alpha+\theta_\beta-\phi)\right),
\end{equation}
and
\begin{equation}
\Cov\left({\hat{J}_z},{\hat{N}}\right)
=\frac{1}{2}\left(
\frac{\sinh^22r}{2}-\frac{\sinh^22z}{2}
+\vert\beta\vert^2\left(\cosh2r-\sinh2r\cos(2\theta_\beta-\theta)\right)
-\vert\alpha\vert^2\left(\cosh2z-\sinh2z\cos\left(2\theta_\alpha-\phi\right)\right)
\right).
\end{equation}
The results for the three considered PMCs \emph{i. e.} Eqs.~\eqref{eq:PMC1_optimal_sqzcoh_sqzcoh}, \eqref{eq:PMC2_optimal_sqzcoh_sqzcoh} and \eqref{eq:PMC3_optimal_sqzcoh_sqzcoh} are detailed in Tab.~\ref{tab:SmyCov_and_Cov_sqzcoh_sqzcoh}.

By setting $\beta=0$ we obtain all the needed results for the squeezed coherent plus squeezed vacuum input discussed in Sec.~\ref{subsec:opt_perf_sqzcoh_sqzvac}. By setting $z=0$, too, we get the needed results for the coherent plus squeezed vacuum input discussed in Sec~\ref{subsec:opt_perf_coh_sqzvac}.

\begin{table}
\begin{tabular}{|c|c|c|c|}
\hline
input 
& Eq.~\eqref{eq:psi_in_sqzcoh_plus_sqzcoh} 
& Eq.~\eqref{eq:psi_in_sqzcoh_plus_sqzcoh} 
& Eq.~\eqref{eq:psi_in_sqzcoh_plus_sqzcoh} \\
state 
& with (PMC1) 
& with (PMC2) 
& with (PMC3) \\
\hline
$\SymCov\left({\hat{J}_x},{\hat{J}_y}\right)$ 
& 0 
& 0 
& 0 \\
\hline
$\SymCov\left({\hat{J}_x},{\hat{J}_z}\right)$  
& $-\frac{\vert\alpha\beta\vert(\sinh2r+\sinh2z)}{4}$ 
& 0 
& 0 \\

\hline
$\SymCov\left({\hat{J}_y},{\hat{J}_z}\right)$   
& 0 
& 0 
& $\frac{\vert\alpha\beta\vert\left(\sinh{r}e^{r}-\sinh{z}e^{z}\right)}{2}$ \\
\hline
$\Cov\left({\hat{J}_x},{\hat{N}}\right)$   
& $\vert\alpha\beta\vert(1-\sinh{r}e^{-r}+\sinh{z}e^{z})$ 
& $\vert\alpha\beta\vert(1-\sinh{r}e^{-r}-\sinh{z}e^{-z})$ 
& 0 \\
\hline
$\Cov\left({\hat{J}_y},{\hat{N}}\right)$   
& 0  
& 0 
& $\vert\alpha\beta\vert(1+\sinh{r}e^{r}+\sinh{z}e^{z})$ \\
\hline
$\Cov\left({\hat{J}_z},{\hat{N}}\right)$  
& $\frac{\sinh^22r-\sinh^22z}{4}+\frac{{\vert\beta\vert^2}e^{-2r}-{\vert\alpha\vert^2}e^{2z}}{2}$
& $\frac{\sinh^22r-\sinh^22z}{4}+\frac{{\vert\beta\vert^2}e^{-2r}-{\vert\alpha\vert^2}e^{-2z}}{2}$ 
& $\frac{\sinh^22r-\sinh^22z}{4}+\frac{{\vert\beta\vert^2}e^{2r}-{\vert\alpha\vert^2}e^{2z}}{2}$ \\
\hline
\end{tabular}
\caption{The covariances and symmetrized covariances for the squeezed-coherent plus squeezed coherent input state given in Eq.~\eqref{eq:psi_in_sqzcoh_plus_sqzcoh}.}
\label{tab:SmyCov_and_Cov_sqzcoh_sqzcoh}
\end{table}


\end{widetext}
\twocolumngrid

%
%
\bibliographystyle{apsrev4-1}

\bibliography{Optimal_phase_sensitivity_unbal_MZI_bibtex}

\end{document}